 \documentclass[twocolumn]{aastex631}

\def\fig{Figure~}
\def\sect{Section~}

\def\sun{$_\odot$}
\def\um{$\mu$m}
\def\h2{H$_2$}
\def\kms{km\,s$^{-1}$}
\def\g34{G34}
\def\hii{\mbox{H{\sc ii}}}

\begin{document}

\title{Detection of an Explosive Outflow in G34.26+0.15}

\author[0000-0002-7881-689X]{Namitha Issac}\thanks{E-mail:namiann@gmail.com}
\affiliation{Shanghai Astronomical Observatory, Chinese Academy of Sciences, 80 Nandan Road, Shanghai 200030, People’s Republic of China \\
}

\author[0000-0003-2619-9305]{Xing Lu}
\affiliation{Shanghai Astronomical Observatory, Chinese Academy of Sciences, 80 Nandan Road, Shanghai 200030, People’s Republic of China \\
}

\author[0000-0002-5286-2564]{Tie Liu}
\affiliation{Shanghai Astronomical Observatory, Chinese Academy of Sciences, 80 Nandan Road, Shanghai 200030, People’s Republic of China \\
}

\author[0000-0003-2343-7937]{Luis A. Zapata}
\affiliation{Instituto de Radioastronom\'{i}a y Astrof\'{i}sica, Universidad Nacional Aut\'{o}noma de M\'{e}xico, 58090, Morelia, Michoac\'{a}n, M\'{e}xico\\
}

\author[0000-0003-3343-9645]{Hong-Li Liu}
\affiliation{School of Physics and Astronomy, Yunnan University, Kunming, 650091, People’s Republic of China \\
}

\author[0000-0001-5917-5751]{Anandmayee Tej}
\affiliation{Indian Institute of Space Science and Technology, Thiruvananthapuram 695 547, Kerala, India
}

\author[0000-0001-7817-1975]{Yan-Kun Zhang}
\affiliation{Shanghai Astronomical Observatory, Chinese Academy of Sciences, 80 Nandan Road, Shanghai 200030, People’s Republic of China \\
}

\author[0000-0001-9822-7817]{Wenyu Jiao}
\affiliation{Shanghai Astronomical Observatory, Chinese Academy of Sciences, 80 Nandan Road, Shanghai 200030, People’s Republic of China \\
}

\author[0000-0002-8389-6695]{Suinan Zhang}
\affiliation{Shanghai Astronomical Observatory, Chinese Academy of Sciences, 80 Nandan Road, Shanghai 200030, People’s Republic of China \\
}



\begin{abstract}
This paper reports on the detection of a likely explosive outflow in the high-mass star-forming complex G34.26+0.15, adding to the small number (six) of explosive outflows detected so far. ALMA CO($2-1$) and SiO($5-4$) archival observations reveal multiple outflow streamers from G34.26+0.15, which correlate well with {\h2} jets identified from \textit{Spitzer}-IRAC 4.5\,{\um} and [4.5]/[3.6] flux ratio maps. These nearly linear outflow streamers originate from a common center within an ultracompact\,{\hii} region located in the complex. The velocity spread of the outflow streamers ranges from 0 to 120\,{\kms}. The radial velocities of these streamers follow the Hubble-Lema\^{i}tre velocity law, indicating an explosive nature. From the CO emission, the total outflow mass, momentum, and outflow energy are estimated to be $\sim$264\,M\sun, 4.3$\times 10 ^3$\,M\sun\,\kms, and 10$^{48}$\,erg, respectively. The event triggering the outflow may have occurred about 19,000 years ago and could also be responsible for powering the expanding UC\,{\hii} region, given the similar dynamical ages and positional coincidence of the UC\,{\hii} region with the origin of the outflow. The magnetic field lines in the region associated with G34.26+0.15 also appear to align with the direction of the outflow streamers and jets, possibly being dragged by the explosive outflow.

\end{abstract}



\section{Introduction} \label{sec:intro}
Highly energetic outflows of explosive nature are a new subclass of molecular outflows detected in massive star-forming regions. Unlike the typical, relatively long-lived collimated bipolar outflows which dissipate the excess angular momentum during the phase of mass accretion onto the forming young stellar objects (YSOs), explosive outflows are impulsive and short-lived and believed to be powered by single, brief energetic events with energy injections of $\sim 10^{47-49}$\,erg \citep{Bally2005}. A sudden ejection of gravitational potential energy, possibly triggered by the dynamical rearrangement of a non-hierarchical massive young stellar system like a stellar merger or by a protostellar collision can drive explosive outflows \citep{{Zapata2009},{Rivilla2014},{Bally2016},{Bally2017}}.
\citet{Zapata2017,Zapata2019} have drawn a clear morphological and kinematic distinction between classical protostellar and explosive outflows.
Explosive outflows are characterized by the presence of several filament-like molecular gas streamers, isotropically distributed in the sky, each of which follows the Hubble-Lema\^{i}tre velocity law where the radial velocities of the filaments increase linearly with the projected distance from the central source. The isotropic distribution makes the red and blueshifted streamers to appear to overlap on the plane of the sky.
The gas streamers are often traced by CO and SiO molecular lines with the emission reaching radial velocities of up to 100\,{\kms}. Aside from the gas streamers, {\h2} ``wakes" and [FeII] ``fingers" can also be observed towards the tips of the filaments as observed in the case of Orion~Becklin–Neugebauer~(BN)/Kleinman–Low~(KL) \citep{Bally2015,Youngblood2016}.
Based on high angular resolution polarization observations, (Orion~BN/KL; \citealt{Cortes2021} and G5.89$–$0.39; \citealt{Fernandez-Lopez2021}), it is also believed that explosive outflows are strong enough to drag the magnetic field lines and rearrange them in a quasi-radial orientation with respect to the origin of the outflow.

In recent years, a few explosive outflow sources were identified from the molecular gas kinematics, namely Orion~BN/KL \citep{Zapata2009}, DR21 \citep{Zapata2013}, G5.89-0.39 \citep{Zapata2019,Zapata2020}, IRAS~16076-5134 \citep{Guzman-Ccolque2022}, Sh2-106 \citep{Bally2022}, and  IRAS~12326-6245 \citep{Zapata2023}. From these discoveries, \citet{Zapata2023} have estimated the rate of explosive outflow events to be one every 90\,yr in the Milky Way. Interestingly, as noted by these authors, this rate is comparable to the approximate rate of supernovae of one in 50\,yr \citep{Diehl2006} and also the massive star formation rate of one in 50\,yr. These indicate that the dynamic interactions in massive young stellar systems like stellar mergers or protostellar collisions might be common occurrences in the initial stages of massive star formation in high-density clustered environments, also leading to the ejection of runaway stars as seen in the case of Orion~BN/KL \citep{Rodriguez2005,Gomez2008,Bally2011}.

In this paper we discuss the likelihood of G34.26+0.15 (hereafter G34) being an explosive outflow source. Located at a distance of 3.3\,kpc \citep{Kuchar1994}, G34 is a high-mass star-forming complex. It consists of four radio components, two of which are hypercompact (HC) {\hii} regions, and the others are a cometary ultracompact (UC) and a shell-like {\hii} regions designated as A, B, C, and D, respectively throughout literature \citep[e.g.,][]{Reid1985,Garay1986,Sewilo2004}.
\citet{Liu2013} suggest that the expansion of the {\hii} region, D is responsible for inducing the sequential star formation in G34, which is also supported in a recent study by \citet{Khan2024}.
The cometary UC\,{\hii} region, C, consists of a ``head" and a diffuse ``tail" that points (from tail through head; see {\fig}\ref{fig:cont_Hii_SiO}a) in the direction of the supernova remnant (SNR) W44 that lies a projected distance of $\sim$40\,pc. However, the SNR shell extends to only $\sim$25\,pc in radius and hence is not yet at a distance close enough to influence the gas motion in G34 \citep{Reid1985}.
This UC\,{\hii} region also harbours a chemically rich hot molecular core (HMC) \citep{MacDonald1995}. \citet{Mookerjea2007} propose that the UC\,{\hii} region is primarily responsible for energizing the HMC which shows no evidence of internal heating. These authors also found the presence of several nitrogen and oxygen bearing complex organic molecules (COMs). \citet{Hajigholi2016} detected ammonia (NH$_3$) lines towards the G34 HMC that shows inverse P-Cygni profiles suggesting mass infall onto the central source. G34 has been catalogued as an ``outflow-only'' candidate by \citet{Cyganowski2008} from the large-scale \textit{Spitzer} Galactic Legacy Infrared Mid-Plane Survey Extraordinaire \citep[GLIMPSE;][]{Benjamin2003}, where multiple jet/outflow like structures are seen in the IRAC 4.5\,{\um} band extending in several directions away from the central object. Such emission may arise from the {\h2}\,($v=0-0$, S(9,10,11)) lines and/or CO\,($v=1-0$) bandheads that are excited by shocks from outflows \citep[e.g.,][]{Noriega-Crespo2004,2006ApJ...645.1264S}. 
According to \citet{Cyganowski2008}, the $[3.6]-[4.5]$ color of the extended emission fall within the range of ``shocked outflow nebulosity'' determined for the DR21 outflow \citep[see {\fig}7 of][]{Smith2006}.
In the near-infrared regime, \citet{Lee2013} have identified several isolated {\h2} knots from the {\h2}$-K$ continuum subtracted image of G34 and classified them as candidate {\h2} outflows.
Furthermore, several SiO outflows have also been detected towards the north-west, south-east, and north-east of G34 \citep{Hatchell2001}.

In presenting the evidence to support the likelihood of G34 being an explosive outflow, we make use of the observations from the archives of ALMA, VLA, and JCMT. The paper is organized as follows. {\sect}\ref{sec:obs} outlines the observations and data reduction details. In {\sect}\ref{sec:results} we discuss the identification of outflows and whether G34 is associated with an explosive outflow event. The conclusions from this study are presented in {\sect}\ref{sec:conclusion}.

\section{Archival Observations} \label{sec:obs}
\subsection{ALMA data}
We make use of the Band 6 Atacama Large Millimeter/submillimeter Array (ALMA) archival data to investigate the gas kinematics of the region associated with G34. The 12-m array observations were carried out on 2019 November 28 using 43 antennas, with baselines ranging from 15 to 313\,m (Project ID: 2019.1.00263.S; PI: John Bally). The observations were made in mosaic mode, consisting of 53 pointings distributed in a Nyqusit-sampled grid with a total on-source time of approximately 23 minutes. The average precipitable water vapor during the observations was 1.1 mm. 
The phase center of the observations was located at the sky position 
$\alpha_{J2000}$=18$^{\rm h}$53$^{\rm m}$15$^{\rm s}$.871 and $\delta_{J2000}$=01$^\circ$15$'$07$''$.131.
The largest recoverable scale of this observation is 11$''$.7. Of the four spectral windows (SPWs), we focus on SPW1 and SPW3 centred at 231.065 and 217.648\,GHz, covering the transitions $^{12}$CO($2-1$) at 230.538\,GHz and SiO($5-4$) at 217.105\,GHz, respectively. We make use of the line-free channels from all the four SPWs to obtain the continuum image. J1924-2914 was used as the flux and bandpass calibrator, while J1851+0035 was used as the phase calibrator.

The data were calibrated and imaged using the Common Astronomy Software Applications (CASA) Version 5.6.1-8. The imaging was done employing the task TCLEAN with the Robust parameter set to +0.5. The continuum map generated has an rms noise level of 0.5\,mJy\,beam$^{-1}$ and an angular resolution of 1$''$.5$\times$1$''$.1, equivalent to spatial resolution of $\sim$0.02\,pc ($\sim$4000\,AU). The CO($2-1$) and SiO($5-4$) cubes have similar angular resolutions of 1$''$.5$\times$1$''$.2 and 1$''$.6$\times$1$''$.3, respectively, and a velocity resolution of 1.5\,{\kms}. 

\subsection{VLA data}
To probe the ionized emission associated with G34, we use the 8.46\,GHz (3.6\,cm) data retrieved from the National Radio Astronomy Observatory VLA (Very Large Array) Archive Survey\footnote{\url{http://www.vla.nrao.edu/ astro/nvas/}} (NVAS). The observation was carried out on 1991 December 6 using the VLA B/A configuration (Legacy ID: AW303; D. Wood). The 3.6\,cm map has an angular resolution of 0$''$.8$\times$0$''$.7 and rms noise level of 0.2\,mJy\,beam$^{-1}$. 

\subsection{JCMT data}
The orientation of magnetic field in the region associated with G34 is derived from the archival 850\,{\um} polarization data (Project ID: M16AD003; PI: Sarah Graves). The observations were done with SCUBA-2/POL-2 \citep{Holland2013,Friberg2016,Friberg2018} mounted on the the James Clerk Maxwell Telescope (JCMT) in the POL-2 DAISY scanning mode. The effective beam is $14.1''$ at 850\,$\mu$m \citep{Dempsey2013}. The data are reduced following the standard procedure\footnote{\url{http://starlink.eao.hawaii.edu/docs/sc22.htx/sc22.html}} for SCUBA-2/POL-2 observations using the STARLINK package SMURF \citep{Chapin2013,Currie2014}. The final \textit{I}, \textit{Q}, and \textit{U} maps are in units of pW. They are converted to the units of Jy\,beam$^{-1}$ by applying a flux correction factor (FCF) of 725\,Jy\,beam$^{-1}$\,pW$^{-1}$. Due to the flux loss from POL-2, value of FCF is 1.35 times larger than for the standard SCUBA-2 FCF of 537\,Jy\,beam$^{-1}$\,pW$^{-1}$ \citep{Dempsey2013}. The rms noise level of the total intensity (Stokes \textit{I}) is measured to be 40\,mJy\,beam$^{-1}$. The polarization angles are derived following the procedure described in \citet{Gu2024}. Only the polarization vectors where the non-polarized intensity ($I$), polarized intensity ($P$), and whose polarization percentage ($p$) satisfying the criteria $I/\delta I \geq 10$, $P/\delta P \geq 3$, and $\delta p \leq 5\%$, respectively were selected. The magnetic field orientation is obtained by rotating the polarization vectors by 90$^\circ$.

\section{Results and Discussions} \label{sec:results}
\subsection{Identification of outflows} \label{sec:outflow}
\begin{figure*}
    \centering
    \includegraphics[scale=0.32]{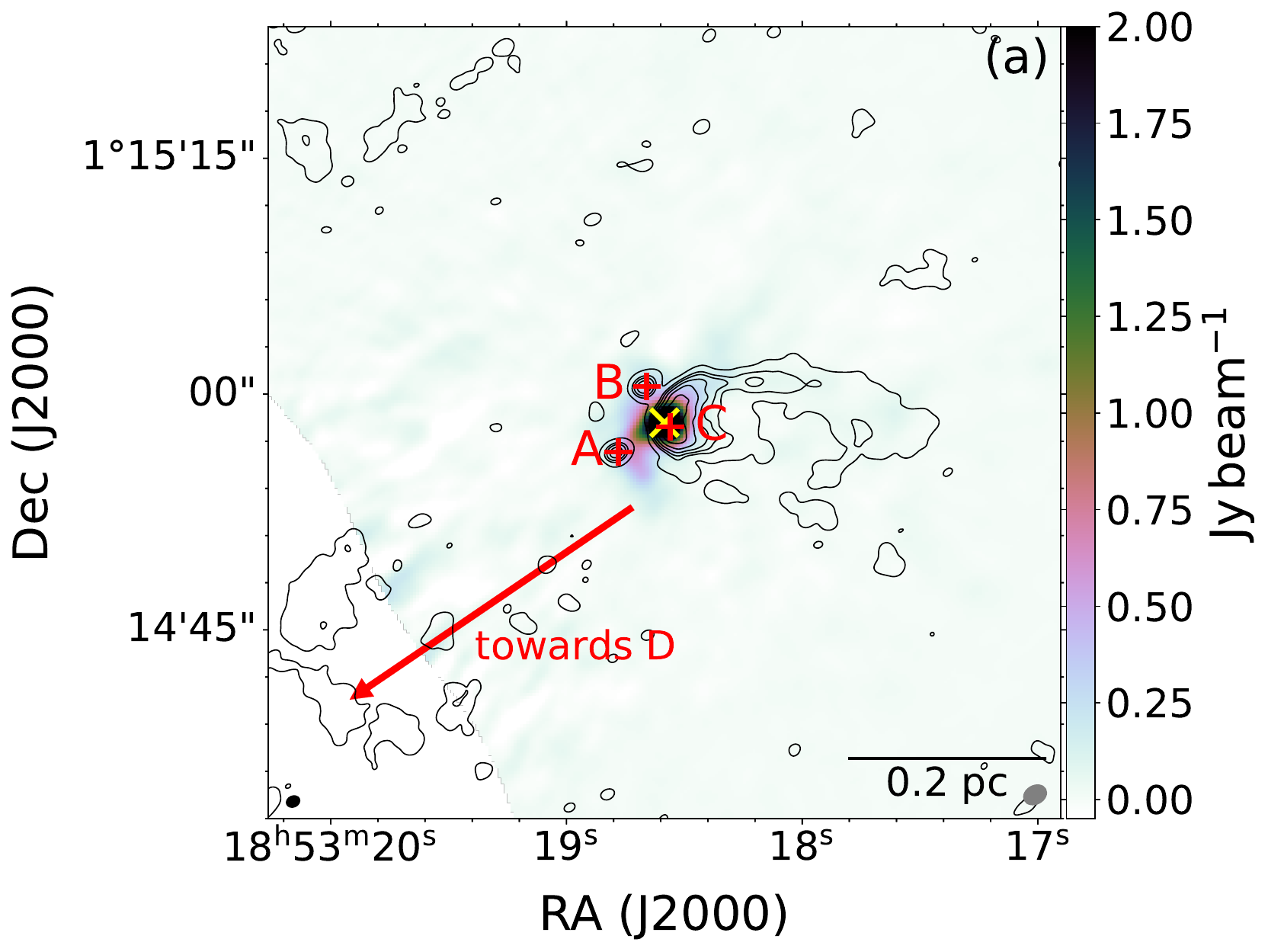} \includegraphics[scale=0.32]{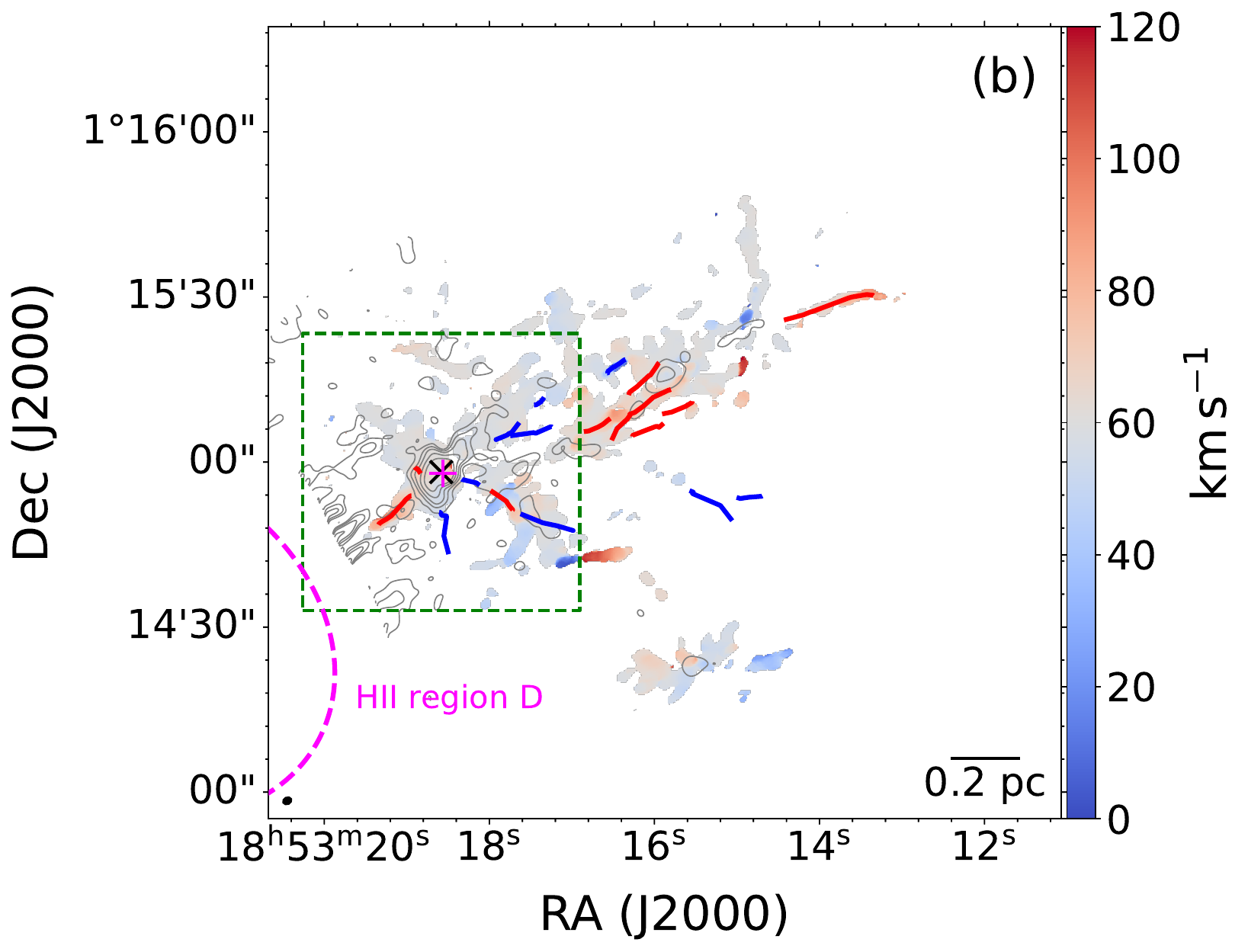} \includegraphics[scale=0.29]{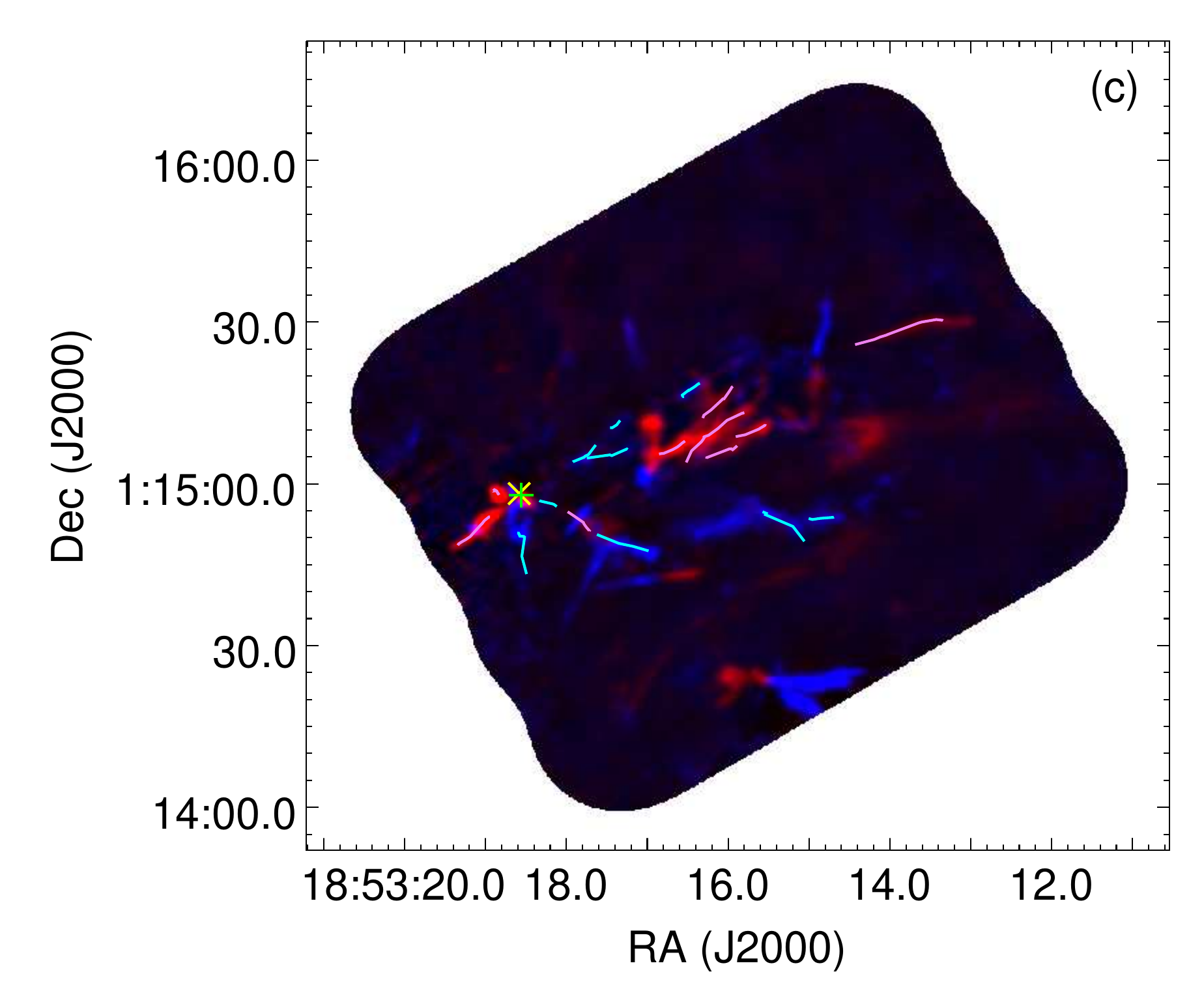}
    \caption{(a) The 1.3\,mm continuum map of G34 is depicted in the colourscale. The peak position of HMC is marked with a `$\times$'. The contours of the 3.6\,cm emission is overlaid in black with contour levels 5, 50, 100, 200, 400, 800, 1200 times $\sigma$ ($\sigma=0.2$\,mJy\,beam$^{-1}$). The two HC\,{\hii} regions, A and B, and the UC\,{\hii} region, C   are indicated by `+'. The arrow indicates the direction towards the shell-like {\hii} region, D. The beams of the 1.3\,mm and 3.6\,cm continuum are shown at the bottom-right and -left corners, respectively. 
    (b) SiO($5-4$) moment-one map of G34 within the velocity range 0 to 120\,{\kms} overlaid with the receding (blue curves) and approaching (red curves) CO($2-1$) outflow streamers identified. The contours of the 1.3\,mm emission is also overlaid in gray with contour levels 10, 80, 160, 320, and 640 times $\sigma$ ($\sigma = 0.5$\,mJy\,beam$^{-1}$). The positions of the HMC and the UC\,{\hii} region are marked with `$\times$' and `+', respectively. A portion of the {\hii} region, D is depicted by the dashed ellipse. The beam of the SiO($5-4$) cube is shown at the bottom-left corner. The green box denotes the field-of-view of (a). (c) Two-colour composite moment-zero maps of CO($2-1$) integrated over the velocity ranges 0 to 48\,{\kms} (blue) and 68 to 120\,{\kms} (red). The red and blueshifted outflow streamers are show in cyan and pink, respectively. 
    }
    \label{fig:cont_Hii_SiO}
\end{figure*}
The ALMA 1.3\,mm continuum map of G34 is shown in {\fig}\ref{fig:cont_Hii_SiO}(a). The peak position of the continuum emission ($\alpha_{J2000}=18^{\rm h}53^{\rm m}18.56^{\rm s}$,$\delta_{J2000}=+01^\circ14'57.90''$) is determined using the 2D Gaussian fitting tool of CASA viewer. From the fitting to the central dense core emission, we obtain an integrated flux density of 9.3$\pm$0.5\,Jy and a peak flux of 4.5$\pm$0.1\,Jy\,beam$^{-1}$. The HMC identified by \citet{Mookerjea2007} at 2.8\,mm coincides with the 1.3\,mm peak. 
The contours of the radio continuum emission at 3.6\,cm is overlaid on this figure. The positions of the two HC\,{\hii} regions (A and B) and the extended cometary UC\,{\hii} region (C) obtained from \citet{Sewilo2004} are marked. The direction towards the shell-like {\hii} region (D) is also labelled. The 1.3\,mm peak coincides with the UC\,{\hii} region, C (within $\sim 0''.5$). 

In {\fig}\ref{fig:cont_Hii_SiO}(b) we present the SiO($5-4$) moment-one (intensity-weighted velocity) map towards G34 within the velocity range 0 to 120\,{\kms}, overlaid with the contours of the 1.3\,mm continuum emission.
The velocity range chosen to construct the SiO($5-4$) moment-one includes radial velocities close to the systemic velocity of the ambient gas of the G34 cloud ($V_{\rm sys}= +58$\,{\kms}; \citealt{Hoang2023}). Since SiO emission primarily traces shocks, the contamination from the ambient gas is considered minimal.
To trace the large-scale outflows around the G34 complex we use the CO($2-1$) data cube, with which we construct channel maps ({\fig}\ref{fig:channel_maps}) having a channel width of 2\,{\kms}. 
Each channel shows several localized emission features. The positions of these features are determined by linearized least-square fits to Gaussian ellipsoids using the task SAD of the Astronomical Image Processing Software (AIPS). 
Examining these features at consecutive velocity channels within the velocity window 0 to 120\,{\kms}, we have discerned 20 filamentary gas streamers with consistent velocity increments and having almost linear structures. The velocity channels in range 48 to 68\,{\kms}, where the emission is dominated by the ambient cloud and has spatially extended structures, are excluded while extracting the outflow streamers. Each streamer traces a sequence of CO($2-1$) condensations. 
Of the outflow streamers identified, 9 are receding (redshifted) reaching radial velocities of up to 62\,{\kms}, and 9 of them are approaching (blushifted) reaching up to $-58$\,{\kms} with respect to 58\,{\kms}, the systemic velocity of G34. Most of these outflow streamers, depicted by red and blue curves in {\fig}\ref{fig:cont_Hii_SiO}(b), nearly follow straight lines. Both the red and blue streamers appear to overlap on the plane of the sky and seem to be radially distributed from a common centre as their origin. 
The SiO($5-4$) emission also reveals filaments tracing the CO streamers and pointing back to the origin of the outflow.

The CO($2-1$) moment-zero map of G34 integrated over the velocity ranges 0 to 48\,{\kms} and 68 to 120\,{\kms} is presented in the two-colour composite image in {\fig}\ref{fig:cont_Hii_SiO}(c). The outflow streamers identified are also marked. The figure shows a few possible bipolar outflows likely originating from low-mass protostars. We have taken care to avoid these outflows while extracting the outflow streamers. 
CO protostellar outflows originating from the HMC have also been detected  from the ``Querying Underlying mechanisms of massive star formation with ALMA-Resolved gas Kinematics and Structures \citep[QUARKS;][]{Liu2024}" survey (K. Huang et al., in prep). However, these outflows are not detected at the resolution of the ALMA data presented in this paper. Additionally, there are no large-scale outflow streamers observed along the direction of the protostellar outflows from the HMC. Thus, it is unlikely that any of the outflow streamers identified are misnomered. 

\begin{figure}
    \centering
    \includegraphics[scale=0.3]{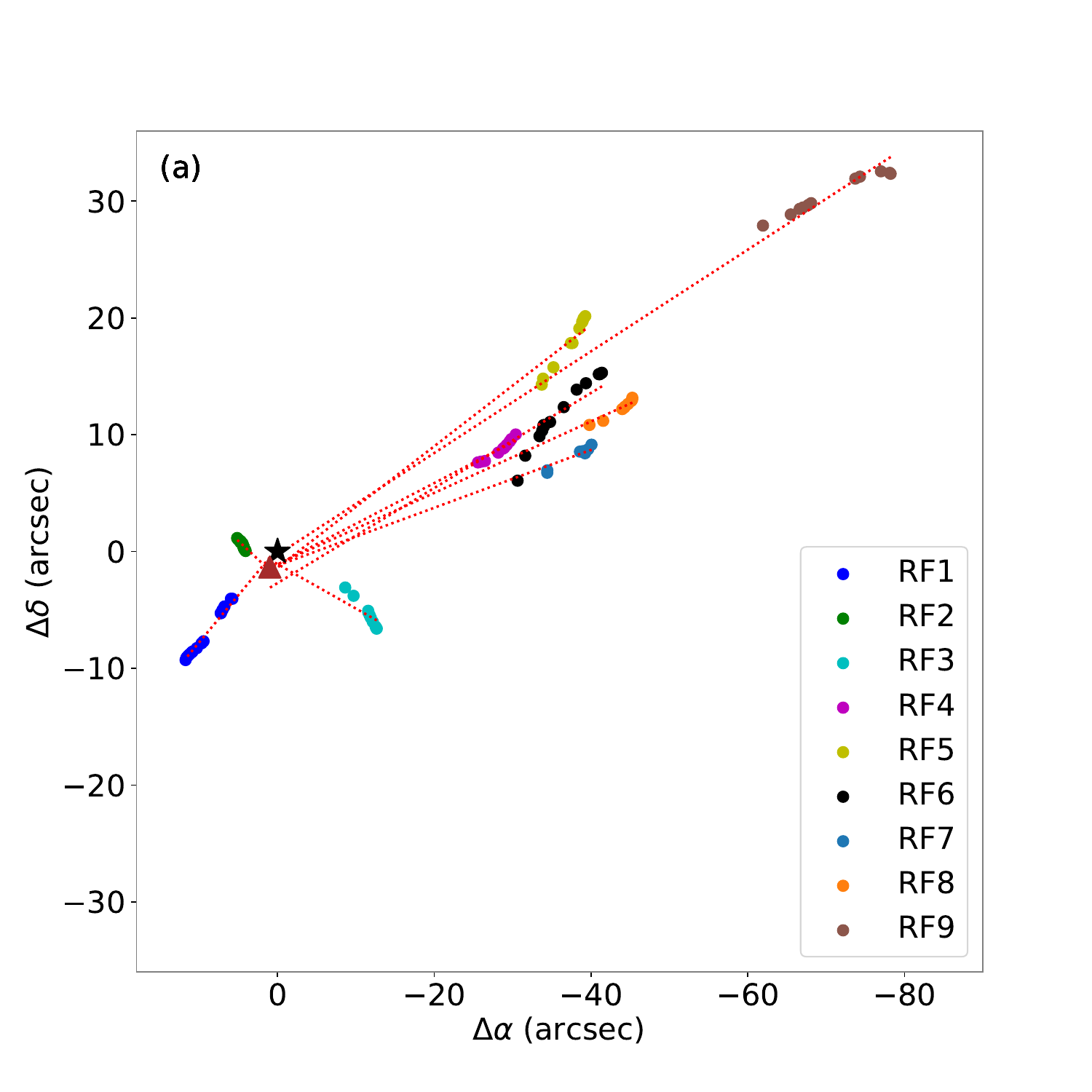} \quad\includegraphics[scale=0.3]{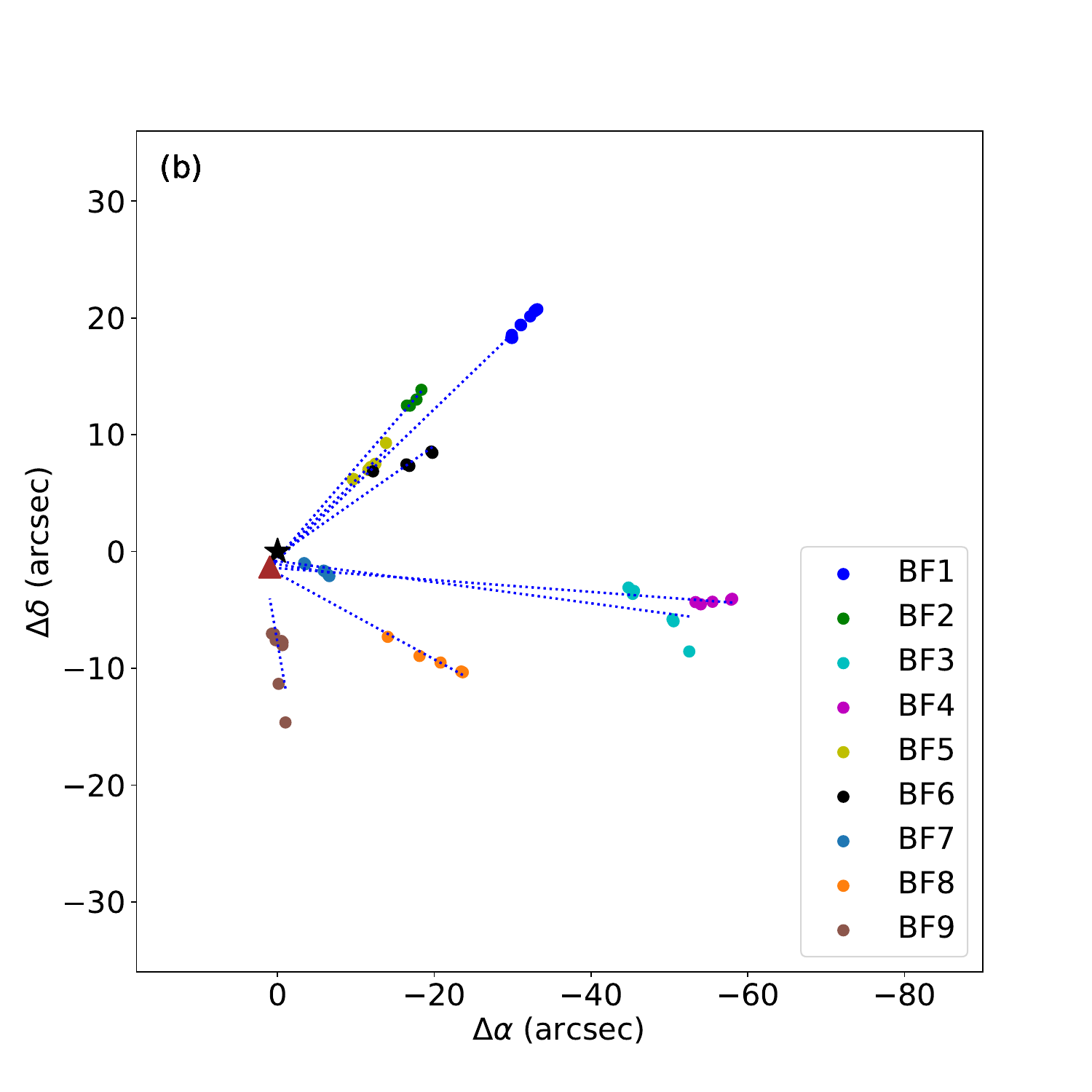}
    \caption{Redshifted (a) and blueshifted (b) CO($2-1$) condensations detected in each CO($2-1$) velocity channel towards G34. The velocity of the  redshifted emission ranges from 68 to 120\,{\kms} and of the blueshifted emission from 0 to 48\,{\kms}. The median position of the intersection points of the outflows which is the possible origin of the outflow is indicated by the brown triangle. The position of the UC\,{\hii} region is denoted by the black star. The red and blue dotted lines in (a) and (b), respectively show the different orientations of the streamers with respect to the origin of the outflow.}
    \label{fig:outflow_RB}
\end{figure}
We plot the CO($2-1$) condensations identified from each velocity channel of the CO($2-1$) cube, with respect to the position of the cometary UC\,{\hii} region, C in {\fig}\ref{fig:outflow_RB}(a) and (b). The red (RF1-RF9) and blueshifted (BF1-BF9) streamers are labeled in this figure. 
Following \citet{Guzman-Ccolque2024}, we find the position of the origin of the outflows. By performing a linear fit on all the CO streamers, we created a dataset of intersection points for each pair of streamers.
Excluding the intersection points more than 5$''$ away from the HMC and the UC\,{\hii} region, we are left with 7 blue streamers (BF1, BF2, BF5, BF6, BF7, BF8, and BF9) and 6 red streamers (RF1, RF2, RF3, RF4, RF7, and RF8). The origin of the outflow is derived by estimating the median position of the intersection points of these 13 streamers and is found to be located at $\alpha_{J2000}$=18$^{\rm h}$53$^{\rm m}$18$^{\rm s}$.63$\pm$0.06${\rm s}$, $\delta_{J2000}$=01$^\circ$14$'$56$''$.56$\pm$0.4$''$. 
This position lies within the UC\,{\hii} region towards the south-east edge at a separation of $\sim 1''.6$ from the peak position of the HMC. The red and blue dotted lines in {\fig}\ref{fig:outflow_RB}, which represent the least-square fits of all the streamers along with the position of the origin, traces the orientation and path of each outflow streamer.

\begin{figure*}
    \centering
    \includegraphics[scale=0.32]{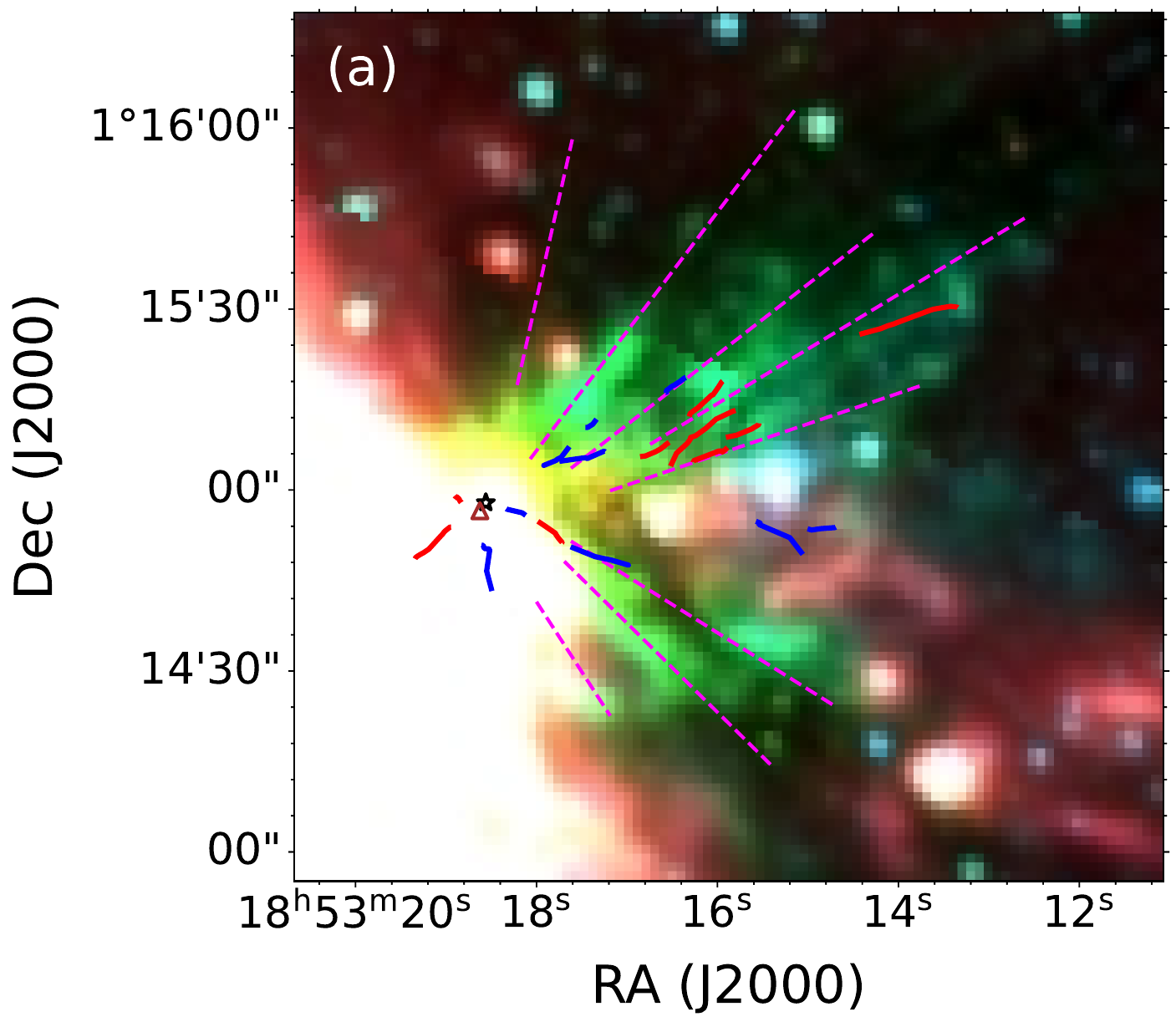} \quad\includegraphics[scale=0.32]{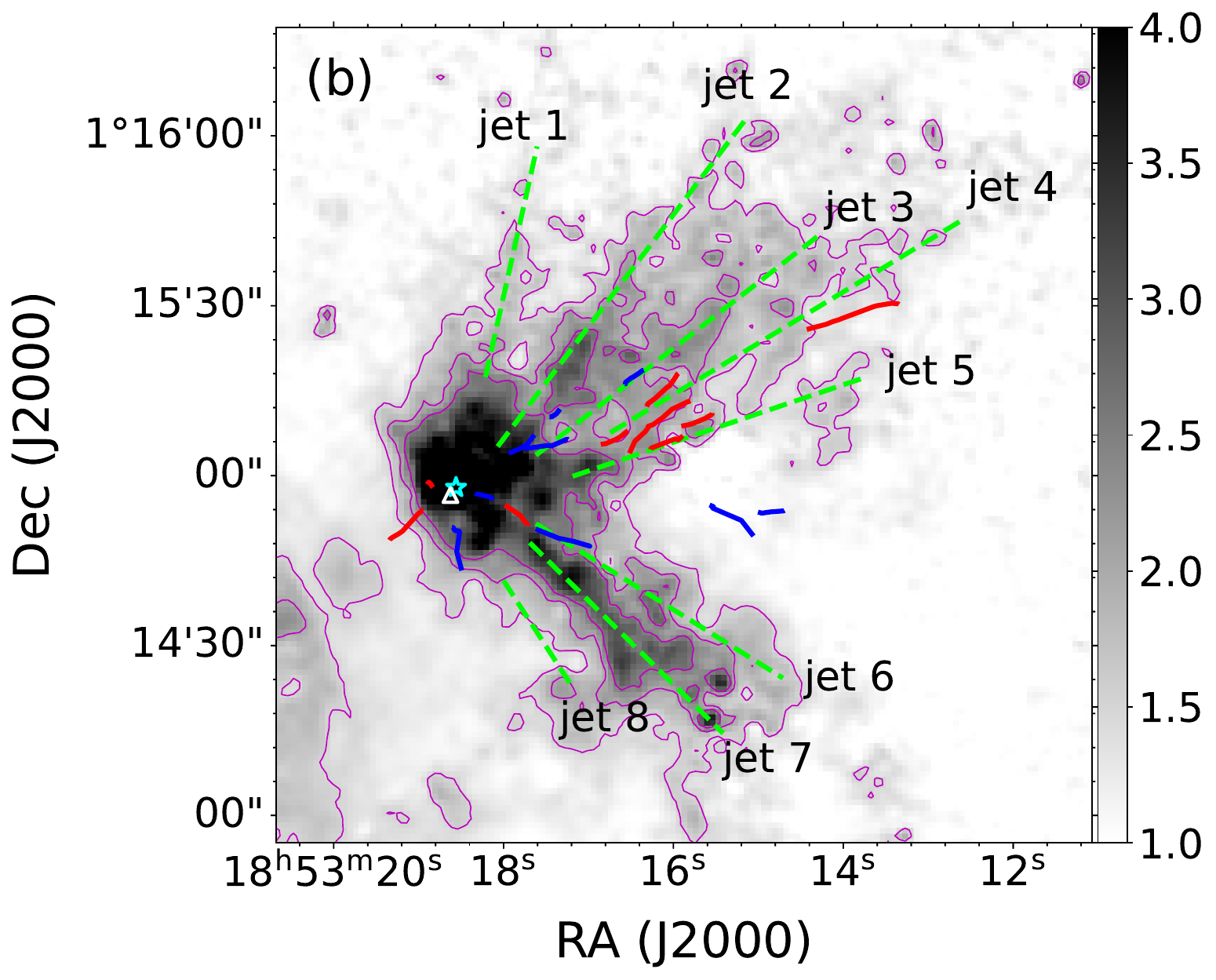}
    \caption{(a) Colour-composite image of G34 from IRAC 3.6\,{\um} (blue), 4.5\,{\um} (green), and 8.0\,{\um} bands. (b) IRAC [4.5]/[3.6] flux ratio map with contour levels 1.5, 2.0, and 2.5. The image is smoothed to 3 pixels to improve the contrast against the background. The {\h2} jets identified from both these images are depicted by the dashed lines. The red and blushifted CO outflow streamers are also overlaid. The hollow star and triangle mark the positions of the UC\,{\hii} region and the possible origin of the outflow, respectively.}
    \label{fig:4.5_outflow}
\end{figure*}
In addition to the molecular outflow traced by the CO($2-1$) and SiO($5-4$) lines, we see evidence of multiple jets in the mid-infrared regime in the G34 region.  
{\fig}\ref{fig:4.5_outflow}(a) is the three color-composite image from the GLIMPSE survey. The emission in the IRAC 4.5\,{\um} (color-coded as green in the IRAC color-composite images) shows several finger-like diverging structures, resembling multiple jets. 
Towards the south-east of the extended 4.5\,{\um} emission in {\fig}\ref{fig:4.5_outflow}(a) is the infrared dust bubble MWP2G0342631+0013065 \citep{Jayasinghe2019} which bounds the shell-like {\hii} region, D \citep{Liu2013}.
\citet{Khan2024} suggest that this bright infrared emission dominated by the 8\,{\um} emission results from the expansion of the {\hii} region that compresses the gas around it, leading to the formation of a shock front.
To reduce the contamination from the stellar components at 4.5\,{\um}, we construct the IRAC [4.5]/[3.6] flux ratio map, presented in {\fig}\ref{fig:4.5_outflow}(b). Along the direction of the jet-like features seen in the 4.5\,{\um} emission, the flux ratio is $\gtrsim$ 1.5, as opposed to stars with flux ratio $\ll$ 1.5 \citep{Takami2010}. Such emission has been interpreted to be tracing the shock-excited 0-0~S(9) line of {\h2} at 4.695\,{\um} \citep{Noriega-Crespo2004}. Comparing the two images, we have visually identified several jets (jet 1 - jet 8), depicted by the dashed lines in both figures that seem to originate from the G34 complex.
The CO outflow streamers are also overlaid on {\fig}\ref{fig:4.5_outflow}. The 4.5\,{\um} jets and the outflow streamers show a good correlation with the jets also roughly pointing towards the origin of the outflow streamers depicted by the triangle in {\fig}\ref{fig:4.5_outflow}.

\subsection{Evidence of explosive outflows} \label{sec:possible outflow}
The $^{12}$CO($2-1$) and infrared observations towards Orion~BN/KL have revealed a massive (10\,M\sun) and energetic ($\sim 10^{47}$\,erg) outflows produced by a violent explosion likely caused by an \textit{N}-body interaction resulting in the ejection of the stars BN, ``source I", and ``source x" about 550\,yr ago \citep{Luhman2017,Bally2020}.
\citet{Zapata2009} suggest that such an isotropic distribution of CO outflows and {\h2} finger-like emission is unlike a typical protostellar outflow seen in star-forming regions. Similar results have been reported for DR21 as well by \citet{Zapata2013} and \citet{Guzman-Ccolque2024}, wherein they propose that the CO and {\h2} emission maybe driven by an explosive event that occurred about 8,600\,yr ago.
Considering the distribution of CO and {\h2} emission around the G34 complex, one can envisage a similar scenario as that of Orion~BN/KL and DR21 in G34 as well. However, the shock front from the expansion of the {\hii} region, D could influence the dynamics of the outflow streamers in G34 and thus affect the isotropy of the outflow distribution expected for an explosive outflow.

\begin{figure*}
    \centering
    \includegraphics[scale=0.28]{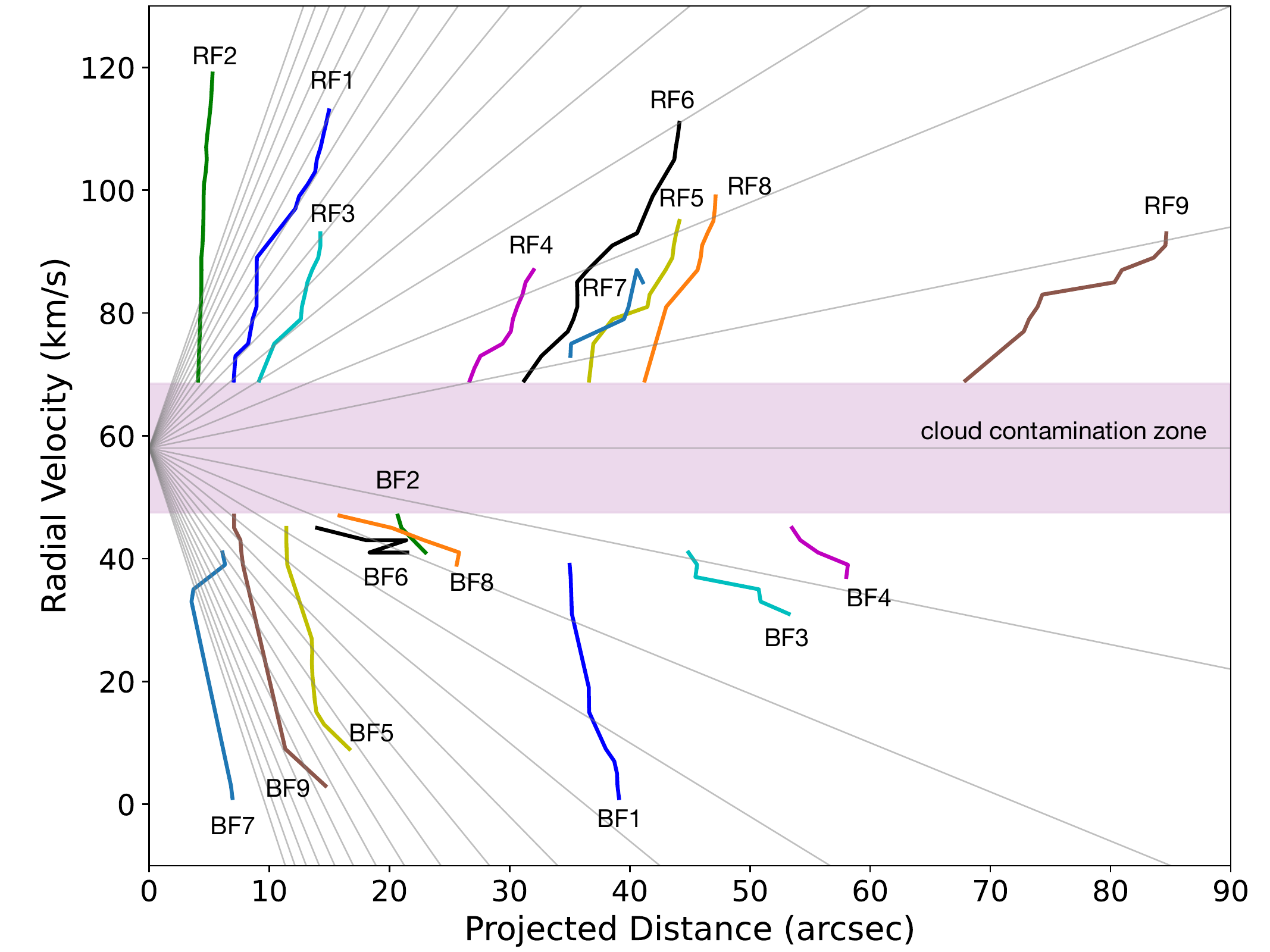}
    \caption{Position-velocity diagram of the outflow streamers identified towards G34. Each streamer identified is designated a different colour and labelled as RF1-RF9 and BF1-BF9 for the red and blueshifted streamers, respectively. The gray lines indicate the linear trend between the projected distance and radial velocity starting from the projected distance of 0$''$ corresponding to the origin of the outflow and  58\,{\kms}, the systemic velocity G34. The shaded portion indicates the region where the emission is contaminated by the ambient cloud.}
    \label{fig:pv}
\end{figure*}
The relation between on-the-sky distance and radial velocity of the 20 identified outflow streamers is plotted in {\fig}\ref{fig:pv}. The gray lines represent the linear trend between the projected distance and the radial velocity predicted by the Hubble-Lema\^{i}tre velocity law. The outflow streamers seem to qualitatively follow this linear trend where the radial velocity of the CO($2-1$) condensations increase with the projected distance from the common centre following the Hubble-Lema\^{i}tre velocity law.
The kinematic behavior of a linear increase in radial velocity with the projected distance is regarded as one of the most distinctive signatures of explosive outflows and have been confirmed in the cases of other explosive outflows \citep[e.g.,][]{Zapata2009,Zapata2013,Zapata2020,Zapata2023,Guzman-Ccolque2024}.
However, the linear trend is not very clear for some of the streamers which have large velocity dispersion (e.g., BF1, RF6, RF8). 
A deviation from the linear trend may be attributed to the shocks originating from the interaction of the outflows with the surrounding material \citep[e.g.,][]{Guzman-Ccolque2024}. The high-mass star-forming region, G34 is classified as a hub-filament system with ongoing mass accretion through filaments \citep{Khan2024}. Some of the filaments lie along the direction of the outflow streamers \citep[refer to {\fig}6 of][]{Khan2024}. The interaction of the outflows with the infalling gas would produce strong shocks resulting in steeper velocity gradients in the position-velocity plot. This scenario is supported by the presence of strong SiO emission, an excellent shock tracer, along the outflow streamers (see {\fig}\ref{fig:cont_Hii_SiO}b).

Assuming local thermodynamical equilibrium and the $^{12}$CO($2-1$) emission to be optically thin at velocities beyond $\pm$10\,\kms from $V_{\rm sys}$, we estimate the mass, momentum, and energy of the explosive outflow following Equations (A1-A4) from \citet{Li2020}.
Taking a distance of 3.3\,kpc, excitation temperature of 29\,K \citep{Urquhart2018} and CO abundance of 10$^{-4}$ \citep{Blake1987}, these values are calculated for the streamers in each channel and then summed to obtain a total mass, momentum, and outflow energy of $\sim$264\,M\sun, 4.3$\times 10^3$\,M\sun\,\kms, and 10$^{48}$\,erg, respectively. This implies that the outflow emission in G34 is associated with a highly energetic event. The estimated outflow energy of 10$^{48}$\,erg is similar to the energies derived for the previously identified explosive outflow events. 
Furthermore, the initial explosion energy would be much larger than the outflow energy because most of the it would have been radiated away by the shocks.

From the emission at 1.3\,cm and 2.8\,mm, \citet{Mookerjea2007} estimated the spectral index of the UC\,{\hii} region to be 0.2, consistent with optically thin emission. Assuming the emission to be optically thin at 3.6\,cm as well, we estimate the Lyman continuum flux, $N_{\rm Ly}$ to be 3.4$\times 10^{48}$\,s$^{-1}$ that translates to an ionizing ZAMS star of spectral type O7-O7.5 \citep{Panagia1973}.
We compute the dynamical age of the UC\,{\hii} region from the 3.6\,cm map using the following expression.
\begin{equation}
    t_{\rm dyn} = \bigg[\frac{4\, R_{\rm s}}{7\, c_{\rm i}}\bigg] \bigg[\bigg(\frac{R_{\hii}}{R_{\rm s}}\bigg)^{7/4} - 1\bigg]
\end{equation}
Here $R_{\rm s} = (3N_{\rm Ly}/4\pi n_0^2 \alpha_{\rm B})^{1/3}$ is the radius of the Str\"omgren sphere, where $n_0 = 1.6 \times 10^5$\,cm$^{-3}$ is the number density of the neutral hydrogen medium. 
Since a large fraction of diffuse emission is lost in the high-resolution 1.3\,mm map due to missing flux effects during interferometric observations and because the 1.3\,mm emission is also contaminated by free-free emission from the UC\,{\hii} region \citep{Liu2013}, we estimate $n_0$ from the single-dish JCMT 850\,{\um} map. 
$\alpha_{\rm B}$ is the radiative recombination coefficient taken to be $2.6\times10^{-13}$\,cm$^3$\,s$^{-1}$ \citep{Kwan1997}.
$R_{\hii} = (A/\pi)^{0.5}$ is the effective radius of the UC\,{\hii} regions, where $A$ is the area within the 3$\sigma$ ($\sigma = 0.5$\,mJy\,beam$^{-1}$) contour of the UC\,{\hii} region.
$c_{\rm i}$ is the isothermal sound speed in the ionized medium, typically assumed to be 10\,{\kms}. The dynamical age of the UC\,{\hii} region is estimated to be $\sim$17,000\,yr. 

We also estimate of the dynamical age of the outflow assuming that all the streamers move with the same velocity. This gives a range of maximum radial velocities for the outflows with varying inclinations relative to the line-of-sight.
Taking the maximum observed radial velocity with respect to the systemic velocity of G34, $\sim$62\,{\kms}, representing the outflow closest to the line-of-sight, and a maximum projected distance of $\sim 74''$ which is the farthest outflow, the dynamical age of the outflow is estimated to be $\sim$19,000\,yr. 
It is to be noted that this gives an order of magnitude estimate at best, considering we have not taken into account the inclination angle of the streamers with the line-of-sight and have assumed an isotropic distribution of the outflow streamers.

The positional coincidence between the centre of the outflow streamers and the UC\,{\hii} region ({\sect}\ref{sec:outflow}) along with their similar dynamical timescales suggest a possible relationship between the expansion of the UC\,{\hii} region and the outflow, where the two were probably triggered by the same explosive event. 
Nonetheless, we cannot dismiss the possibility that the UC\,{\hii} region predates the explosive event, given the large uncertainty in the age estimation of the explosive outflow.

Similar expanding {\hii} regions are also seen within the explosive outflows in DR21 \citep{Zapata2013,Guzman-Ccolque2024}, G5.89$-$0.39 \citep{Zapata2019,Zapata2020}, and IRAS~12326$-$6245 \citep{Zapata2023} where the expanding shell is centred at the origin of the outflows. 
These studies suggest that the explosive event that drives the outflows is also responsible for powering the expanding {\hii} regions. 
Currently there are no known massive young stellar objects (MYSO) at the peak of the UC\,{\hii} region in G34. It is possible that the protostar may have been ejected during the explosive event, as in the case of Orion BN/KL \citep{Bally2011,Rodriguez2020} and DR21 \citep{Zapata2013,Guzman-Ccolque2024}. An in-depth proper motion study would be required to confirm this. 
Furthermore, several compact hot cores are detected along the periphery of the cometary head of the UC\,{\hii} region (K. Huang et al., in prep) which could indicate triggered star formation post explosion.

All the characteristics including the Hubble-Lema\^{i}tre-like expansion motion of the streamers and high outflow energy suggest that G34 is a likely explosive outflow candidate.

\subsection{Outflow-dragged magnetic field} \label{sec:magnetic field}
\begin{figure}
    \centering
    \includegraphics[scale=0.32]{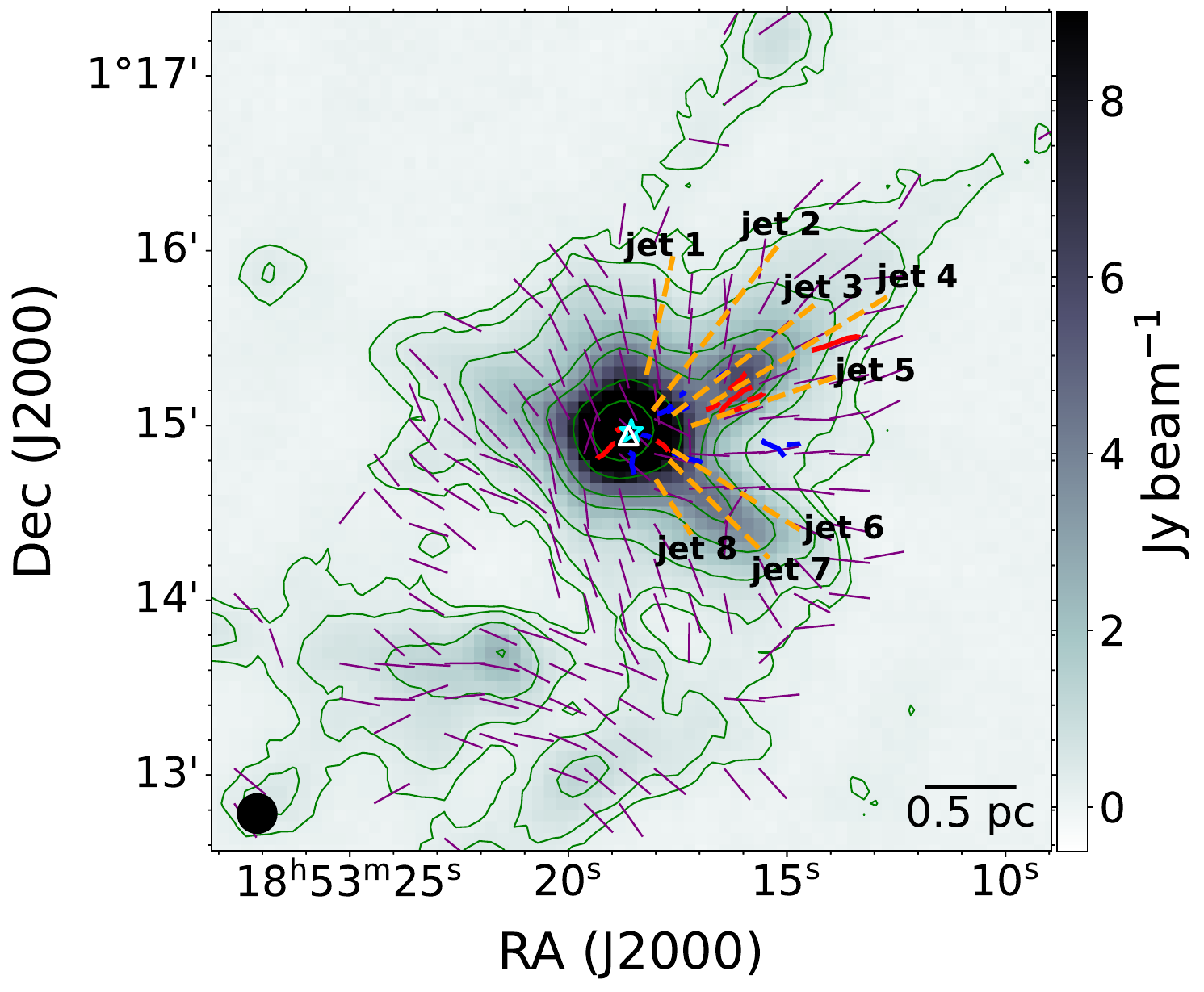}
    \caption{The 850\,{\um} Stokes \textit{I} map of the region surrounding G34 from JCMT SCUBA-2/POL-2 with contour levels 5, 10, 25, 65, 100, 200, and 500 $\sigma$ ($\sigma=0.04$\,Jy\,beam$^{-1}$). The purple line segments represent the magnetic field direction as inferred from the polarization data. The red and blushifted CO outflow streamers (red and blue curves) and the {\h2} jets identified from {\fig}\ref{fig:4.5_outflow} (dashed lines) are also overlaid. The hollow star and triangle mark the positions of the UC\,{\hii} region and the possible origin of the outflow, respectively. The beam is shown on the bottom-left corner.}
    \label{fig:850_mag}
\end{figure}
In {\fig}\ref{fig:850_mag} we present the 850\,{\um} Stokes \textit{I} map from the SCUBA-2/POL-2 archival observations towards G34. The magnetic field lines inferred from the polarization data are overlaid on this figure, along with the CO outflow streamers and {\h2} jets. Even with the coarse resolution of the single-dish polarization observations, one can see that the magnetic field is approximately oriented along the direction of the outflows streamers and jets. 
To quantify the relative alignment of the magnetic field lines to the {\h2} jets, we calculate the angle difference between the directions of the two (see Appendix~\ref{sec:Bfield_jet}). {\fig}\ref{fig:850_jet} shows a plot of the same, where we see that the angle difference for most magnetic field lines and {\h2} jets fall within $-20^\circ$ to $20^\circ$. It is to be noted most of the magnetic field lines along jet 1, jet 2, and jet 6 have angle difference $> |20^\circ|$. Nonetheless, the median value of the angle difference between the magnetic field lines and the {\h2} jets is $\sim -5^\circ$. This prompts us to infer that the explosive outflow could be responsible for aligning the magnetic field along the outflow direction.

Of the six explosive outflow sources identified, magnetic fields have been reported towards Orion~BN/KL \citep{Cortes2021} and G5.89$–$0.39 \citep{Fernandez-Lopez2021} from high angular resolution dust polarization observations using ALMA.
From the 1.3 and 3.1\,mm dust polarization observations \citet{Cortes2021} found the magnetic field to be orientated quasi-radially within $\sim$5000\,au from the origin of the explosive outflow. The outflows carve cavities in the dust resulting in the polarized dust emission to have an anti-correlation with the outflow streamers and the magnetic field is aligned in the direction of the cavities.
Evaluating the energy balance, they estimate that the explosive outflow may be energetic enough to propel a shock from the centre of the Orion~BN/KL nebula which can drag the magnetic field lines and rearrange them in a quasi-radial orientation in the inner radius of the outflow ($\sim$5000\,au). In the case of G5.89$–$0.39, \citet{Fernandez-Lopez2021} have found that the magnetic field follows a radial pattern towards the `Central Shell' at 1.2\,mm which coincides with the shell-like UC\,{\hii} region, similar to Orion~BN/KL. These authors suggest that such a radial distribution of magnetic field could be a signpost of explosive outflow events. These studies lend support to our inference that the magnetic field lines in G34 are also dragged by the explosive outflow event.
While the magnetic field may be dragged by the explosive outflow, we cannot rule out the possibility that the magnetic field dominates the orientation of the outflows in this region. However, to obtain a better insight requires evaluating the energy balance between the magnetic field and the outflow energies \citep[e.g.][]{Cortes2021}.
This advocates for higher angular resolution polarization observations to enable a reliable estimate of the magnetic field energy.

\subsection{Rate of explosive events in the Milky Way}
From the six explosive outflows reported in literature (Orion~BN/KL, DR21, G5.89-0.39, IRAS~16076-5134, Sh2-106, and  IRAS~12326-6245), \citet{Zapata2023} have estimated the rate of explosive events to be one in every 90\,yr in the Milky Way. 
These authors have also noted that this rate is comparable to the approximate rate of supernovae of one in 50\,yr \citep{Diehl2006} which is also similar to the massive star formation rate of one in 50\,yr.
With the new detection of an explosive outflow in G34, we update the rate of events in our Galaxy following the same method described in \citet{Zapata2023}. Assuming that the explosive events are evenly spaced over a time span of 31,560 yr (the time period covering all the seven outflows and taking into consideration their different distances to earth) and are distributed within a projected circle of radius 2.8\,kpc (the separation between IRAS~16076-5134 and DR21, the farthest separated outflow sources), we extrapolate the frequency of occurrence of explosive events to the disk of our Galaxy which is taken to be a thin disk with a radius of 15\,kpc. This gives a total number of 200 explosive events in the Galaxy and the rate of explosive events to be one in every 160\,yr.
Our estimate is higher than that reported by \citet{Zapata2023}, owing to the larger dynamical age of the explosive outflow in G34 compared to the other six explosive outflows detected so far.
This, however, is a crude estimate based on several assumptions, including the size of the Galaxy and an approximate dynamical age of the outflow. The rate of occurrence explosive events in the Milky Way can be refined as more explosive outflows are detected, and hence, can draw a better correlation with the rate of supernovae and massive star formation.

\section{Conclusions} \label{sec:conclusion}
The ALMA CO($2-1$) and SiO($5-4$) archival observations have revealed the presence of multiple outflow streamers associated with the high-mass star-forming complex, G34. Along with molecular outflows at mm wavelengths, several {\h2} jets have also been identified from the IRAC 4.5\,{\um} and [4.5]/[3.6] flux ratio maps showing a good correlation with the CO outflow streamers. The molecular outflow streamers have nearly linear structures and seem to emanate from a common centre within the UC\,{\hii} region. The radial velocity of each streamer follows the Hubble-Lema\^{i}tre velocity law, indicative of the explosive nature of the outflow.
The explosive event that initiated the outflow appears to have occurred about 19,000\,yr ago. This event may also be the mechanism that powers the expanding UC\,{\hii}, as indicated by their similar dynamical ages and the positional coincidence of the UC\,{\hii} region with the origin of the outflow. Additionally, these explosive outflows might be responsible for aligning the magnetic field along the outflow direction.
Our results add to the small sample of rare explosive outflows observed in our Galaxy.

\begin{acknowledgments}
We thank Prof. John Bally for critically going through the manuscript and giving valuable suggestions. This work is supported by the National Key R\&D Program of China (No.~2022YFA1603101) and the Strategic Priority Research Program of the Chinese Academy of Sciences (CAS) Grant No.~XDB0800300. N.I. acknowledges the support by the China Postdoctoral Science Foundation through grant No. 2023M733624 and the Shanghai Postdoctoral Excellence Program through grant No.~2023682. X.L. acknowledges support from the National Natural Science Foundation of China (NSFC) through grant Nos.~12273090 and 12322305, the Natural Science Foundation of Shanghai (No.~23ZR1482100), and the CAS ``Light of West China'' Program No.\ xbzg-zdsys-202212. L.A.Z. acknowledges financial support from CONACyT-280775, UNAM-PAPIIT IN110618, and IN112323 grants, México. H.-L.L is supported by Yunnan Fundamental Research Project 202401AS070121, and by Xingdian Talent Support Plan–Youth Project.
This paper makes use of the following ALMA data: ADS/JAO.ALMA\#2019.1.00263.S. ALMA is a partnership of ESO (representing its member states), NSF (USA) and NINS (Japan), together with NRC (Canada), MOST and ASIAA (Taiwan), and KASI (Republic of Korea), in cooperation with the Republic of Chile. The Joint ALMA Observatory is operated by ESO, AUI/NRAO and NAOJ. This paper makes use of the data obtained with JCMT. The JCMT is operated by the East Asian Observatory on behalf of the National Astronomical Observatory of Japan, the Academia Sinica Institute of Astronomy and Astrophysics, the Korea Astronomy and Space Science Institute and Center for Astronomical Mega-Science (as well as the National Key Research and Development Program of China with No. 2017YFA0402700). Additional funding support is provided by the Science and Technology Facilities Council of the United Kingdom and participating universities in the United Kingdom and Canada. 
\end{acknowledgments}

%



\software{CASA \citep{McMullin2007,CASA2022}, APLpy \citep{2012ascl.soft08017R}, Astropy \citep{Astropy2013}}



\appendix
\restartappendixnumbering
\section{Channel maps} \label{sec:channel_map}
\begin{figure*}
    \centering
    \includegraphics[scale=0.1]{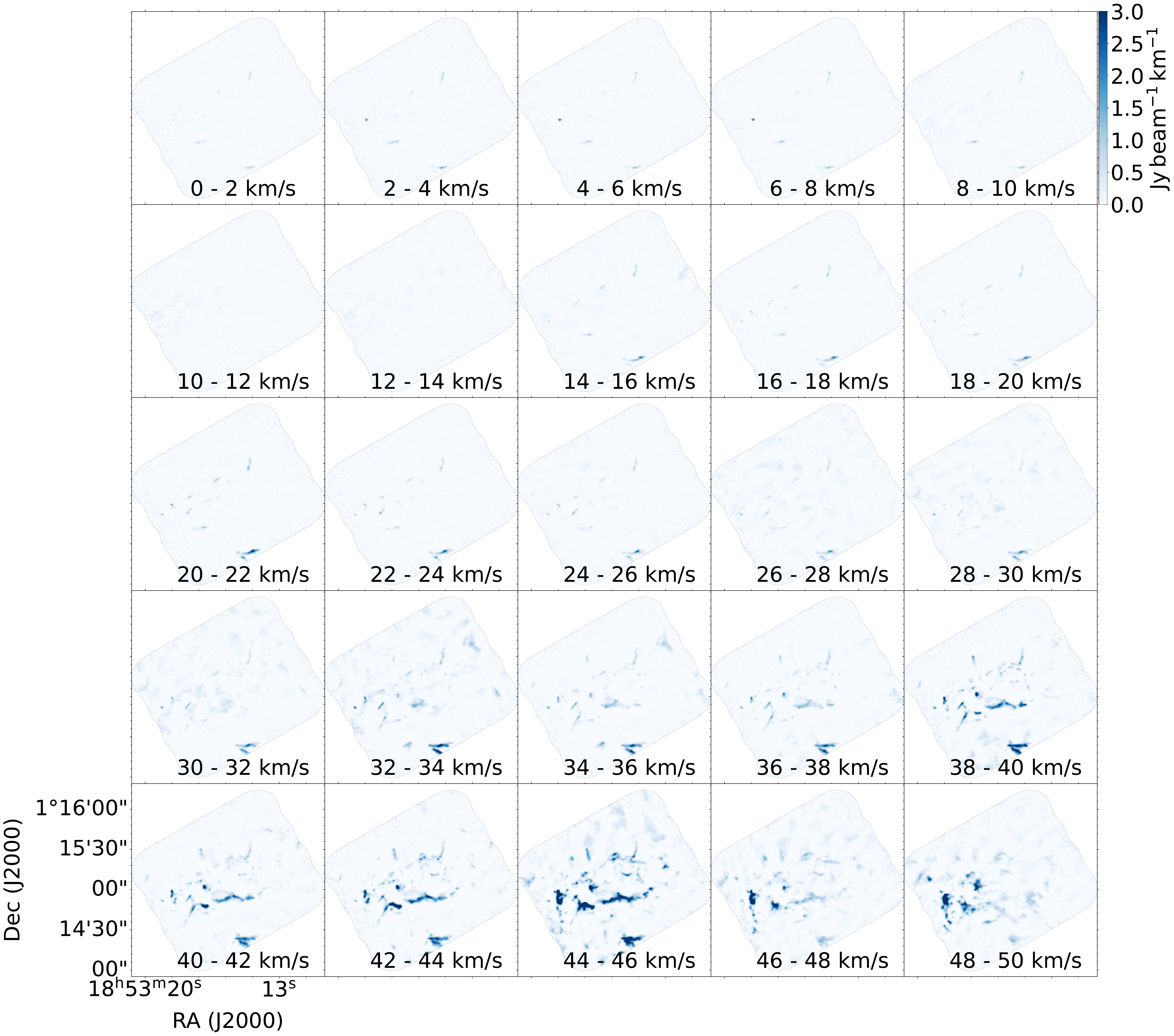}
    \quad\includegraphics[scale=0.1]{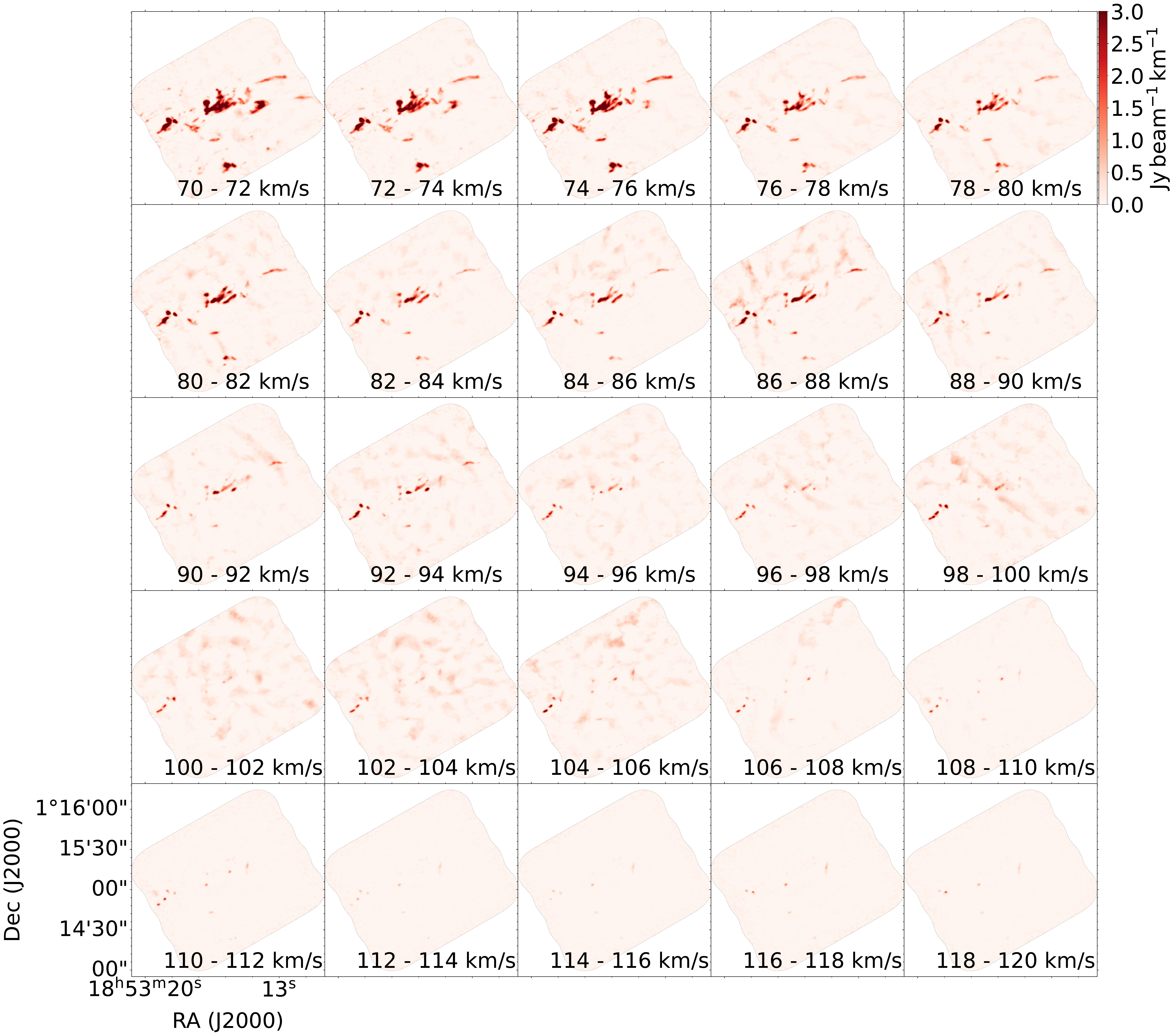}
    \caption{The ALMA $^{12}$CO($2-1$) channel maps from in the velocity range 0 to 50\,{\kms} (left) and 70 to 120\,{\kms} (right) towards G34. The CO condensations in each channel having a nearly linear structure with velocity increment are identified as the outflow streamers. }
    \label{fig:channel_maps}
\end{figure*}

The CO($2-1$) channel maps of width 2\,{\kms} in the velocity range 0 to 50\,{\kms} and 70 to 120\,{\kms} are shown in {\fig}\ref{fig:channel_maps}(a) and (b), respectively. 

\section{Direction of magnetic field lines v/s {\h2} jets}
\label{sec:Bfield_jet}
\begin{figure*}
    \centering
    \includegraphics[scale=0.45]{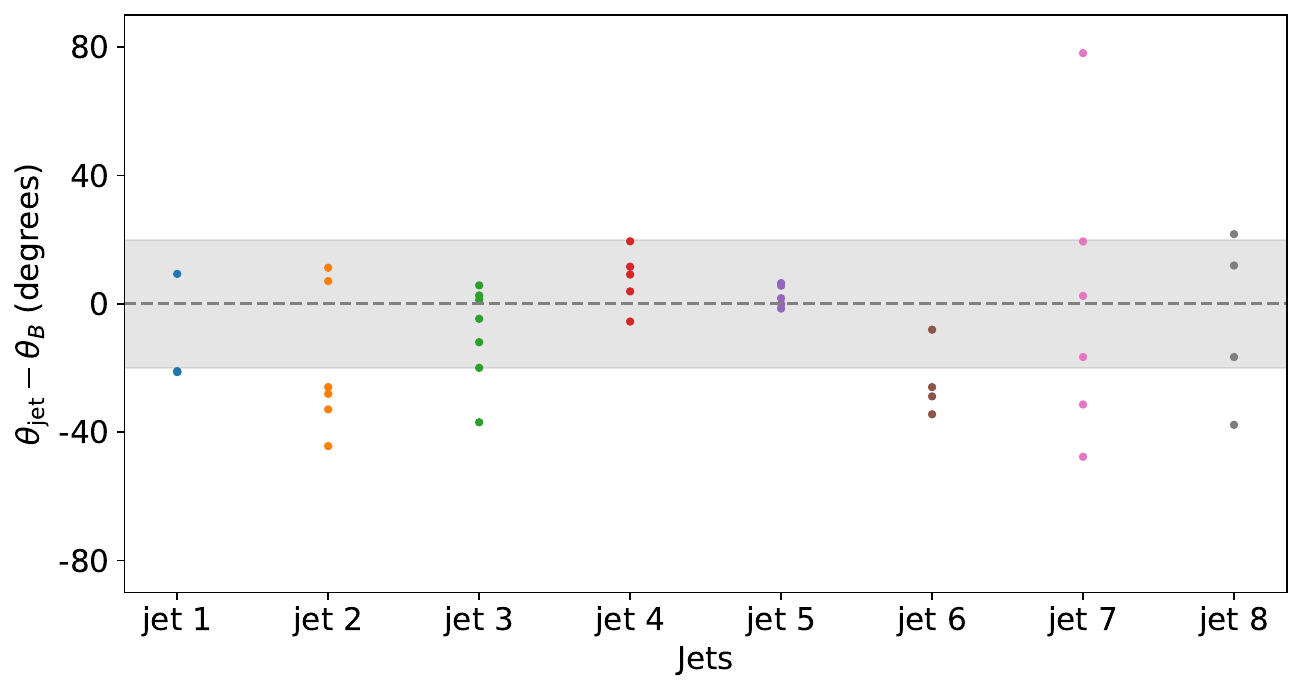}
    \caption{The angle difference between of the magnetic field lines ($\theta_B$) and the {\h2} jet direction ($\theta_{\rm jet}$) along each {\h2} jet identified in G34.}
    \label{fig:850_jet}
\end{figure*}
To quantitatively evaluate the alignment of the magnetic field lines with the {\h2} jets, we compute the angle difference between the two. We select the magnetic field lines along each {\h2} jet and determine the difference in angle between the direction of each magnetic field line ($\theta_B$) and the direction of the corresponding {\h2} jet ($\theta_{\rm jet}$). This is repeated for each {\h2} jet and the resulting difference in orientation between the magnetic field lines and each {\h2} jet is plotted in {\fig}\ref{fig:850_jet}.


\bibliography{reference}{}

\begin{thebibliography}{}
\expandafter\ifx\csname natexlab\endcsname\relax\def\natexlab#1{#1}\fi
\providecommand{\url}[1]{\href{#1}{#1}}
\providecommand{\dodoi}[1]{doi:~\href{http://doi.org/#1}{\nolinkurl{#1}}}
\providecommand{\doeprint}[1]{\href{http://ascl.net/#1}{\nolinkurl{http://ascl.net/#1}}}
\providecommand{\doarXiv}[1]{\href{https://arxiv.org/abs/#1}{\nolinkurl{https://arxiv.org/abs/#1}}}

\bibitem[{{Astropy Collaboration} {et~al.}(2013){Astropy Collaboration}, {Robitaille}, {Tollerud}, {Greenfield}, {Droettboom}, {Bray}, {Aldcroft}, {Davis}, {Ginsburg}, {Price-Whelan}, {Kerzendorf}, {Conley}, {Crighton}, {Barbary}, {Muna}, {Ferguson}, {Grollier}, {Parikh}, {Nair}, {Unther}, {Deil}, {Woillez}, {Conseil}, {Kramer}, {Turner}, {Singer}, {Fox}, {Weaver}, {Zabalza}, {Edwards}, {Azalee Bostroem}, {Burke}, {Casey}, {Crawford}, {Dencheva}, {Ely}, {Jenness}, {Labrie}, {Lim}, {Pierfederici}, {Pontzen}, {Ptak}, {Refsdal}, {Servillat}, \& {Streicher}}]{Astropy2013}
{Astropy Collaboration}, {Robitaille}, T.~P., {Tollerud}, E.~J., {et~al.} 2013, \aap, 558, A33, \dodoi{10.1051/0004-6361/201322068}

\bibitem[{{Bally}(2016)}]{Bally2016}
{Bally}, J. 2016, \araa, 54, 491, \dodoi{10.1146/annurev-astro-081915-023341}

\bibitem[{{Bally} {et~al.}(2022){Bally}, {Chia}, {Ginsburg}, {Reipurth}, {Tanaka}, {Zinnecker}, \& {Faulhaber}}]{Bally2022}
{Bally}, J., {Chia}, Z., {Ginsburg}, A., {et~al.} 2022, \apj, 924, 50, \dodoi{10.3847/1538-4357/ac30de}

\bibitem[{{Bally} {et~al.}(2011){Bally}, {Cunningham}, {Moeckel}, {Burton}, {Smith}, {Frank}, \& {Nordlund}}]{Bally2011}
{Bally}, J., {Cunningham}, N.~J., {Moeckel}, N., {et~al.} 2011, \apj, 727, 113, \dodoi{10.1088/0004-637X/727/2/113}

\bibitem[{{Bally} {et~al.}(2017){Bally}, {Ginsburg}, {Arce}, {Eisner}, {Youngblood}, {Zapata}, \& {Zinnecker}}]{Bally2017}
{Bally}, J., {Ginsburg}, A., {Arce}, H., {et~al.} 2017, \apj, 837, 60, \dodoi{10.3847/1538-4357/aa5c8b}

\bibitem[{{Bally} {et~al.}(2020){Bally}, {Ginsburg}, {Forbrich}, \& {Vargas-Gonz{\'a}lez}}]{Bally2020}
{Bally}, J., {Ginsburg}, A., {Forbrich}, J., \& {Vargas-Gonz{\'a}lez}, J. 2020, \apj, 889, 178, \dodoi{10.3847/1538-4357/ab65f2}

\bibitem[{{Bally} {et~al.}(2015){Bally}, {Ginsburg}, {Silvia}, \& {Youngblood}}]{Bally2015}
{Bally}, J., {Ginsburg}, A., {Silvia}, D., \& {Youngblood}, A. 2015, \aap, 579, A130, \dodoi{10.1051/0004-6361/201425073}

\bibitem[{{Bally} \& {Zinnecker}(2005)}]{Bally2005}
{Bally}, J., \& {Zinnecker}, H. 2005, \aj, 129, 2281, \dodoi{10.1086/429098}

\bibitem[{{Benjamin} {et~al.}(2003){Benjamin}, {Churchwell}, {Babler}, {Bania}, {Clemens}, {Cohen}, {Dickey}, {Indebetouw}, {Jackson}, {Kobulnicky}, {Lazarian}, {Marston}, {Mathis}, {Meade}, {Seager}, {Stolovy}, {Watson}, {Whitney}, {Wolff}, \& {Wolfire}}]{Benjamin2003}
{Benjamin}, R.~A., {Churchwell}, E., {Babler}, B.~L., {et~al.} 2003, \pasp, 115, 953, \dodoi{10.1086/376696}

\bibitem[{{Blake} {et~al.}(1987){Blake}, {Sutton}, {Masson}, \& {Phillips}}]{Blake1987}
{Blake}, G.~A., {Sutton}, E.~C., {Masson}, C.~R., \& {Phillips}, T.~G. 1987, \apj, 315, 621, \dodoi{10.1086/165165}

\bibitem[{{CASA Team} {et~al.}(2022){CASA Team}, {Bean}, {Bhatnagar}, {Castro}, {Donovan Meyer}, {Emonts}, {Garcia}, {Garwood}, {Golap}, {Gonzalez Villalba}, {Harris}, {Hayashi}, {Hoskins}, {Hsieh}, {Jagannathan}, {Kawasaki}, {Keimpema}, {Kettenis}, {Lopez}, {Marvil}, {Masters}, {McNichols}, {Mehringer}, {Miel}, {Moellenbrock}, {Montesino}, {Nakazato}, {Ott}, {Petry}, {Pokorny}, {Raba}, {Rau}, {Schiebel}, {Schweighart}, {Sekhar}, {Shimada}, {Small}, {Steeb}, {Sugimoto}, {Suoranta}, {Tsutsumi}, {van Bemmel}, {Verkouter}, {Wells}, {Xiong}, {Szomoru}, {Griffith}, {Glendenning}, \& {Kern}}]{CASA2022}
{CASA Team}, {Bean}, B., {Bhatnagar}, S., {et~al.} 2022, \pasp, 134, 114501, \dodoi{10.1088/1538-3873/ac9642}

\bibitem[{{Chapin} {et~al.}(2013){Chapin}, {Berry}, {Gibb}, {Jenness}, {Scott}, {Tilanus}, {Economou}, \& {Holland}}]{Chapin2013}
{Chapin}, E.~L., {Berry}, D.~S., {Gibb}, A.~G., {et~al.} 2013, \mnras, 430, 2545, \dodoi{10.1093/mnras/stt052}

\bibitem[{{Cortes} {et~al.}(2021){Cortes}, {Le Gouellec}, {Hull}, {Girart}, {Louvet}, {Fomalont}, {Kameno}, {Moellenbrock}, {Nagai}, {Nakanishi}, \& {Villard}}]{Cortes2021}
{Cortes}, P.~C., {Le Gouellec}, V. J.~M., {Hull}, C. L.~H., {et~al.} 2021, \apj, 907, 94, \dodoi{10.3847/1538-4357/abcafb}

\bibitem[{{Currie} {et~al.}(2014){Currie}, {Berry}, {Jenness}, {Gibb}, {Bell}, \& {Draper}}]{Currie2014}
{Currie}, M.~J., {Berry}, D.~S., {Jenness}, T., {et~al.} 2014, in Astronomical Society of the Pacific Conference Series, Vol. 485, Astronomical Data Analysis Software and Systems XXIII, ed. N.~{Manset} \& P.~{Forshay}, 391

\bibitem[{{Cyganowski} {et~al.}(2008){Cyganowski}, {Whitney}, {Holden}, {Braden}, {Brogan}, {Churchwell}, {Indebetouw}, {Watson}, {Babler}, {Benjamin}, {Gomez}, {Meade}, {Povich}, {Robitaille}, \& {Watson}}]{Cyganowski2008}
{Cyganowski}, C.~J., {Whitney}, B.~A., {Holden}, E., {et~al.} 2008, \aj, 136, 2391, \dodoi{10.1088/0004-6256/136/6/2391}

\bibitem[{{Dempsey} {et~al.}(2013){Dempsey}, {Friberg}, {Jenness}, {Tilanus}, {Thomas}, {Holland}, {Bintley}, {Berry}, {Chapin}, {Chrysostomou}, {Davis}, {Gibb}, {Parsons}, \& {Robson}}]{Dempsey2013}
{Dempsey}, J.~T., {Friberg}, P., {Jenness}, T., {et~al.} 2013, \mnras, 430, 2534, \dodoi{10.1093/mnras/stt090}

\bibitem[{{Diehl} {et~al.}(2006){Diehl}, {Halloin}, {Kretschmer}, {Lichti}, {Sch{\"o}nfelder}, {Strong}, {von Kienlin}, {Wang}, {Jean}, {Kn{\"o}dlseder}, {Roques}, {Weidenspointner}, {Schanne}, {Hartmann}, {Winkler}, \& {Wunderer}}]{Diehl2006}
{Diehl}, R., {Halloin}, H., {Kretschmer}, K., {et~al.} 2006, \nat, 439, 45, \dodoi{10.1038/nature04364}

\bibitem[{{Fern{\'a}ndez-L{\'o}pez} {et~al.}(2021){Fern{\'a}ndez-L{\'o}pez}, {Sanhueza}, {Zapata}, {Stephens}, {Hull}, {Zhang}, {Girart}, {Koch}, {Cort{\'e}s}, {Silva}, {Tatematsu}, {Nakamura}, {Guzm{\'a}n}, {Nguyen Luong}, {Guzm{\'a}n Ccolque}, {Tang}, \& {Chen}}]{Fernandez-Lopez2021}
{Fern{\'a}ndez-L{\'o}pez}, M., {Sanhueza}, P., {Zapata}, L.~A., {et~al.} 2021, \apj, 913, 29, \dodoi{10.3847/1538-4357/abf2b6}

\bibitem[{{Friberg} {et~al.}(2016){Friberg}, {Bastien}, {Berry}, {Savini}, {Graves}, \& {Pattle}}]{Friberg2016}
{Friberg}, P., {Bastien}, P., {Berry}, D., {et~al.} 2016, in Proc. SPIE, Vol. 9914, , 991403, \dodoi{10.1117/12.2231943}

\bibitem[{{Friberg} {et~al.}(2018){Friberg}, {Berry}, {Savini}, {Bintley}, {Dempsey}, {Graves}, \& {Parsons}}]{Friberg2018}
{Friberg}, P., {Berry}, D., {Savini}, G., {et~al.} 2018, in Proc. SPIE, Vol. 10708, , 107083M, \dodoi{10.1117/12.2314345}

\bibitem[{{Garay} {et~al.}(1986){Garay}, {Rodriguez}, \& {van Gorkom}}]{Garay1986}
{Garay}, G., {Rodriguez}, L.~F., \& {van Gorkom}, J.~H. 1986, \apj, 309, 553, \dodoi{10.1086/164624}

\bibitem[{{G{\'o}mez} {et~al.}(2008){G{\'o}mez}, {Rodr{\'\i}guez}, {Loinard}, {Lizano}, {Allen}, {Poveda}, \& {Menten}}]{Gomez2008}
{G{\'o}mez}, L., {Rodr{\'\i}guez}, L.~F., {Loinard}, L., {et~al.} 2008, \apj, 685, 333, \dodoi{10.1086/590229}

\bibitem[{{Gu} {et~al.}(2024){Gu}, {Liu}, {Li}, {Shen}, {Liu}, {Liu}, {Lu}, {Montillaud}, {Jiao}, {Juvela}, {Rawlings}, {Zhang}, {Koch}, {Ristorcelli}, {Carriere}, {Eden}, {Ren}, {Tatematsu}, {Hirano}, {Luo}, {Mai}, \& {Issac}}]{Gu2024}
{Gu}, Q.-L., {Liu}, T., {Li}, P.~S., {et~al.} 2024, \apj, 963, 126, \dodoi{10.3847/1538-4357/ad1bc7}

\bibitem[{{Guzm{\'a}n Ccolque} {et~al.}(2024){Guzm{\'a}n Ccolque}, {Fern{\'a}ndez L{\'o}pez}, {Zapata}, {Bally}, \& {Rivera-Ortiz}}]{Guzman-Ccolque2024}
{Guzm{\'a}n Ccolque}, E., {Fern{\'a}ndez L{\'o}pez}, M., {Zapata}, L.~A., {Bally}, J., \& {Rivera-Ortiz}, P.~R. 2024, \aap, 689, A339, \dodoi{10.1051/0004-6361/202449874}

\bibitem[{{Guzm{\'a}n Ccolque} {et~al.}(2022){Guzm{\'a}n Ccolque}, {Fern{\'a}ndez-L{\'o}pez}, {Zapata}, \& {Baug}}]{Guzman-Ccolque2022}
{Guzm{\'a}n Ccolque}, E., {Fern{\'a}ndez-L{\'o}pez}, M., {Zapata}, L.~A., \& {Baug}, T. 2022, \apj, 937, 51, \dodoi{10.3847/1538-4357/ac8c35}

\bibitem[{{Hajigholi} {et~al.}(2016){Hajigholi}, {Persson}, {Wirstr{\"o}m}, {Black}, {Bergman}, {Olofsson}, {Olberg}, {Wyrowski}, {Coutens}, {Hjalmarson}, \& {Menten}}]{Hajigholi2016}
{Hajigholi}, M., {Persson}, C.~M., {Wirstr{\"o}m}, E.~S., {et~al.} 2016, \aap, 585, A158, \dodoi{10.1051/0004-6361/201526451}

\bibitem[{{Hatchell} {et~al.}(2001){Hatchell}, {Fuller}, \& {Millar}}]{Hatchell2001}
{Hatchell}, J., {Fuller}, G.~A., \& {Millar}, T.~J. 2001, \aap, 372, 281, \dodoi{10.1051/0004-6361:20010468}

\bibitem[{{Hoang} {et~al.}(2023){Hoang}, {Karska}, {Lee}, {Wyrowski}, {Tram}, {Yang}, \& {Menten}}]{Hoang2023}
{Hoang}, T.~D., {Karska}, A., {Lee}, M.~Y., {et~al.} 2023, \aap, 679, A121, \dodoi{10.1051/0004-6361/202347163}

\bibitem[{{Holland} {et~al.}(2013){Holland}, {Bintley}, {Chapin}, {Chrysostomou}, {Davis}, {Dempsey}, {Duncan}, {Fich}, {Friberg}, {Halpern}, {Irwin}, {Jenness}, {Kelly}, {MacIntosh}, {Robson}, {Scott}, {Ade}, {Atad-Ettedgui}, {Berry}, {Craig}, {Gao}, {Gibb}, {Hilton}, {Hollister}, {Kycia}, {Lunney}, {McGregor}, {Montgomery}, {Parkes}, {Tilanus}, {Ullom}, {Walther}, {Walton}, {Woodcraft}, {Amiri}, {Atkinson}, {Burger}, {Chuter}, {Coulson}, {Doriese}, {Dunare}, {Economou}, {Niemack}, {Parsons}, {Reintsema}, {Sibthorpe}, {Smail}, {Sudiwala}, \& {Thomas}}]{Holland2013}
{Holland}, W.~S., {Bintley}, D., {Chapin}, E.~L., {et~al.} 2013, \mnras, 430, 2513, \dodoi{10.1093/mnras/sts612}

\bibitem[{{Jayasinghe} {et~al.}(2019){Jayasinghe}, {Dixon}, {Povich}, {Binder}, {Velasco}, {Lepore}, {Xu}, {Offner}, {Kobulnicky}, {Anderson}, {Kendrew}, \& {Simpson}}]{Jayasinghe2019}
{Jayasinghe}, T., {Dixon}, D., {Povich}, M.~S., {et~al.} 2019, \mnras, 488, 1141, \dodoi{10.1093/mnras/stz1738}

\bibitem[{{Khan} {et~al.}(2024){Khan}, {Pattle}, \& {Graves}}]{Khan2024}
{Khan}, Z.~A., {Pattle}, K., \& {Graves}, S.~F. 2024, \mnras, 535, 107, \dodoi{10.1093/mnras/stae2350}

\bibitem[{{Kuchar} \& {Bania}(1994)}]{Kuchar1994}
{Kuchar}, T.~A., \& {Bania}, T.~M. 1994, \apj, 436, 117, \dodoi{10.1086/174886}

\bibitem[{{Kwan}(1997)}]{Kwan1997}
{Kwan}, J. 1997, \apj, 489, 284, \dodoi{10.1086/304773}

\bibitem[{{Lee} {et~al.}(2013){Lee}, {Liao}, {Froebrich}, {Karr}, {Ioannidis}, {Lee}, {Su}, {Liu}, {Duan}, \& {Takami}}]{Lee2013}
{Lee}, H.-T., {Liao}, W.-T., {Froebrich}, D., {et~al.} 2013, \apjs, 208, 23, \dodoi{10.1088/0067-0049/208/2/23}

\bibitem[{{Li} {et~al.}(2020){Li}, {Sanhueza}, {Zhang}, {Nakamura}, {Lu}, {Wang}, {Liu}, {Tatematsu}, {Jackson}, {Silva}, {Guzm{\'a}n}, {Sakai}, {Izumi}, {Tafoya}, {Li}, {Contreras}, {Morii}, \& {Kim}}]{Li2020}
{Li}, S., {Sanhueza}, P., {Zhang}, Q., {et~al.} 2020, \apj, 903, 119, \dodoi{10.3847/1538-4357/abb81f}

\bibitem[{{Liu} {et~al.}(2013){Liu}, {Wu}, \& {Zhang}}]{Liu2013}
{Liu}, T., {Wu}, Y., \& {Zhang}, H. 2013, \apj, 776, 29, \dodoi{10.1088/0004-637X/776/1/29}

\bibitem[{{Liu} {et~al.}(2024){Liu}, {Liu}, {Zhu}, {Garay}, {Liu}, {Goldsmith}, {Evans}, {Kim}, {Liu}, {Xu}, {Lu}, {Tej}, {Mai}, {Bronfman}, {Li}, {Mardones}, {Stutz}, {Tatematsu}, {Wang}, {Zhang}, {Qin}, {Zhou}, {Luo}, {Zhang}, {Cheng}, {He}, {Gu}, {Li}, {Zhang}, {Zhang}, {Saha}, {Dewangan}, {Sanhueza}, \& {Shen}}]{Liu2024}
{Liu}, X., {Liu}, T., {Zhu}, L., {et~al.} 2024, Research in Astronomy and Astrophysics, 24, 025009, \dodoi{10.1088/1674-4527/ad0d5c}

\bibitem[{{Luhman} {et~al.}(2017){Luhman}, {Robberto}, {Tan}, {Andersen}, {Giulia Ubeira Gabellini}, {Manara}, {Platais}, \& {Ubeda}}]{Luhman2017}
{Luhman}, K.~L., {Robberto}, M., {Tan}, J.~C., {et~al.} 2017, \apjl, 838, L3, \dodoi{10.3847/2041-8213/aa5ff6}

\bibitem[{{MacDonald} {et~al.}(1995){MacDonald}, {Habing}, \& {Millar}}]{MacDonald1995}
{MacDonald}, G.~H., {Habing}, R.~J., \& {Millar}, T.~J. 1995, \apss, 224, 177, \dodoi{10.1007/BF00667841}

\bibitem[{{McMullin} {et~al.}(2007){McMullin}, {Waters}, {Schiebel}, {Young}, \& {Golap}}]{McMullin2007}
{McMullin}, J.~P., {Waters}, B., {Schiebel}, D., {Young}, W., \& {Golap}, K. 2007, in Astronomical Society of the Pacific Conference Series, Vol. 376, Astronomical Data Analysis Software and Systems XVI, ed. R.~A. {Shaw}, F.~{Hill}, \& D.~J. {Bell}, 127

\bibitem[{{Mookerjea} {et~al.}(2007){Mookerjea}, {Casper}, {Mundy}, \& {Looney}}]{Mookerjea2007}
{Mookerjea}, B., {Casper}, E., {Mundy}, L.~G., \& {Looney}, L.~W. 2007, \apj, 659, 447, \dodoi{10.1086/512095}

\bibitem[{{Noriega-Crespo} {et~al.}(2004){Noriega-Crespo}, {Morris}, {Marleau}, {Carey}, {Boogert}, {van Dishoeck}, {Evans}, {Keene}, {Muzerolle}, {Stapelfeldt}, {Pontoppidan}, {Lowrance}, {Allen}, \& {Bourke}}]{Noriega-Crespo2004}
{Noriega-Crespo}, A., {Morris}, P., {Marleau}, F.~R., {et~al.} 2004, \apjs, 154, 352, \dodoi{10.1086/422819}

\bibitem[{{Panagia}(1973)}]{Panagia1973}
{Panagia}, N. 1973, \aj, 78, 929, \dodoi{10.1086/111498}

\bibitem[{{Reid} \& {Ho}(1985)}]{Reid1985}
{Reid}, M.~J., \& {Ho}, P.~T.~P. 1985, \apjl, 288, L17, \dodoi{10.1086/184412}

\bibitem[{{Rivilla} {et~al.}(2014){Rivilla}, {Jim{\'e}nez-Serra}, {Mart{\'\i}n-Pintado}, \& {Sanz-Forcada}}]{Rivilla2014}
{Rivilla}, V.~M., {Jim{\'e}nez-Serra}, I., {Mart{\'\i}n-Pintado}, J., \& {Sanz-Forcada}, J. 2014, \mnras, 437, 1561, \dodoi{10.1093/mnras/stt1989}

\bibitem[{{Robitaille} \& {Bressert}(2012)}]{2012ascl.soft08017R}
{Robitaille}, T., \& {Bressert}, E. 2012, {APLpy: Astronomical Plotting Library in Python}, Astrophysics Source Code Library, record ascl:1208.017

\bibitem[{{Rodr{\'\i}guez} {et~al.}(2020){Rodr{\'\i}guez}, {Dzib}, {Zapata}, {Lizano}, {Loinard}, {Menten}, \& {G{\'o}mez}}]{Rodriguez2020}
{Rodr{\'\i}guez}, L.~F., {Dzib}, S.~A., {Zapata}, L., {et~al.} 2020, \apj, 892, 82, \dodoi{10.3847/1538-4357/ab7816}

\bibitem[{{Rodr{\'\i}guez} {et~al.}(2005){Rodr{\'\i}guez}, {Poveda}, {Lizano}, \& {Allen}}]{Rodriguez2005}
{Rodr{\'\i}guez}, L.~F., {Poveda}, A., {Lizano}, S., \& {Allen}, C. 2005, \apjl, 627, L65, \dodoi{10.1086/432052}

\bibitem[{{Sewilo} {et~al.}(2004){Sewilo}, {Churchwell}, {Kurtz}, {Goss}, \& {Hofner}}]{Sewilo2004}
{Sewilo}, M., {Churchwell}, E., {Kurtz}, S., {Goss}, W.~M., \& {Hofner}, P. 2004, \apj, 605, 285, \dodoi{10.1086/382268}

\bibitem[{{Smith} {et~al.}(2006{\natexlab{a}}){Smith}, {Hora}, {Marengo}, \& {Pipher}}]{2006ApJ...645.1264S}
{Smith}, H.~A., {Hora}, J.~L., {Marengo}, M., \& {Pipher}, J.~L. 2006{\natexlab{a}}, \apj, 645, 1264, \dodoi{10.1086/504370}

\bibitem[{{Smith} {et~al.}(2006{\natexlab{b}}){Smith}, {Hora}, {Marengo}, \& {Pipher}}]{Smith2006}
---. 2006{\natexlab{b}}, \apj, 645, 1264, \dodoi{10.1086/504370}

\bibitem[{{Takami} {et~al.}(2010){Takami}, {Karr}, {Koh}, {Chen}, \& {Lee}}]{Takami2010}
{Takami}, M., {Karr}, J.~L., {Koh}, H., {Chen}, H.-H., \& {Lee}, H.-T. 2010, \apj, 720, 155, \dodoi{10.1088/0004-637X/720/1/155}

\bibitem[{{Urquhart} {et~al.}(2018){Urquhart}, {K{\"o}nig}, {Giannetti}, {Leurini}, {Moore}, {Eden}, {Pillai}, {Thompson}, {Braiding}, {Burton}, {Csengeri}, {Dempsey}, {Figura}, {Froebrich}, {Menten}, {Schuller}, {Smith}, \& {Wyrowski}}]{Urquhart2018}
{Urquhart}, J.~S., {K{\"o}nig}, C., {Giannetti}, A., {et~al.} 2018, \mnras, 473, 1059, \dodoi{10.1093/mnras/stx2258}

\bibitem[{{Youngblood} {et~al.}(2016){Youngblood}, {Ginsburg}, \& {Bally}}]{Youngblood2016}
{Youngblood}, A., {Ginsburg}, A., \& {Bally}, J. 2016, \aj, 151, 173, \dodoi{10.3847/0004-6256/151/6/173}

\bibitem[{{Zapata} {et~al.}(2009){Zapata}, {Schmid-Burgk}, {Ho}, {Rodr{\'\i}guez}, \& {Menten}}]{Zapata2009}
{Zapata}, L.~A., {Schmid-Burgk}, J., {Ho}, P. T.~P., {Rodr{\'\i}guez}, L.~F., \& {Menten}, K.~M. 2009, \apjl, 704, L45, \dodoi{10.1088/0004-637X/704/1/L45}

\bibitem[{{Zapata} {et~al.}(2013){Zapata}, {Schmid-Burgk}, {P{\'e}rez-Goytia}, {Ho}, {Rodr{\'\i}guez}, {Loinard}, \& {Cruz-Gonz{\'a}lez}}]{Zapata2013}
{Zapata}, L.~A., {Schmid-Burgk}, J., {P{\'e}rez-Goytia}, N., {et~al.} 2013, \apjl, 765, L29, \dodoi{10.1088/2041-8205/765/2/L29}

\bibitem[{{Zapata} {et~al.}(2017){Zapata}, {Schmid-Burgk}, {Rodr{\'\i}guez}, {Palau}, \& {Loinard}}]{Zapata2017}
{Zapata}, L.~A., {Schmid-Burgk}, J., {Rodr{\'\i}guez}, L.~F., {Palau}, A., \& {Loinard}, L. 2017, \apj, 836, 133, \dodoi{10.3847/1538-4357/aa5b94}

\bibitem[{{Zapata} {et~al.}(2019){Zapata}, {Ho}, {Guzm{\'a}n Ccolque}, {Fern{\'a}ndez-Lop{\'e}z}, {Rodr{\'\i}guez}, {Bally}, {Sanhueza}, {Palau}, \& {Saito}}]{Zapata2019}
{Zapata}, L.~A., {Ho}, P. T.~P., {Guzm{\'a}n Ccolque}, E., {et~al.} 2019, \mnras, 486, L15, \dodoi{10.1093/mnrasl/slz051}

\bibitem[{{Zapata} {et~al.}(2020){Zapata}, {Ho}, {Fern{\'a}ndez-L{\'o}pez}, {Ccolque}, {Rodr{\'\i}guez}, {Reyes-Vald{\'e}s}, {Bally}, {Palau}, {Saito}, {Sanhueza}, {Rivera-Ortiz}, \& {Rodr{\'\i}guez-Gonz{\'a}lez}}]{Zapata2020}
{Zapata}, L.~A., {Ho}, P. T.~P., {Fern{\'a}ndez-L{\'o}pez}, M., {et~al.} 2020, \apjl, 902, L47, \dodoi{10.3847/2041-8213/abbd3f}

\bibitem[{{Zapata} {et~al.}(2023){Zapata}, {Fern{\'a}ndez-L{\'o}pez}, {Leurini}, {Guzm{\'a}n Ccolque}, {Skretas}, {Rodr{\'\i}guez}, {Palau}, {Menten}, \& {Wyrowski}}]{Zapata2023}
{Zapata}, L.~A., {Fern{\'a}ndez-L{\'o}pez}, M., {Leurini}, S., {et~al.} 2023, \apjl, 956, L35, \dodoi{10.3847/2041-8213/acfe71}

\end{thebibliography}
\bibliographystyle{aasjournal}



\end{document}